*De novo* design of α-helical peptide amphiphiles repairing fragmented collagen type I *via* supramolecular co-assembly

Shanshan Su[1,2#], Jie Yang[1,2#], Guo Zhang[1,2], Zhiquan Yu[1,2], Yuxuan Chen[1,2], Alexander van Teijlingen[3], Dawen Yu[1,2], Tong Li[1,2], Yubin Ke[4], Hua Yang[4], Haoran Zhang[5], Jialong Chen[1,2], Jiaming Sun[1,2*] & Yuanhao Wu[1,2*]

[1]Plastic and Reconstructive Surgery Department, Wuhan Union Hospital, Tongji Medical College, Huazhong University of Science and Technology, Wuhan, China.

[2]Plastic and Reconstructive Surgery Research Institute, Wuhan Union Hospital, Tongji Medical College, Huazhong University of Science and Technology, Wuhan, China.

[3]ModelMole, Glasgow, United Kingdom.

[4]Spallation Neutron Source Science Center, Dalang, Dongguan, China.

[5]Department of Infectious Diseases, Tongji Hospital, Tongji Medical College and State Key Laboratory for Diagnosis and Treatment of Severe Zoonotic Infectious Disease, Huazhong University of Science and Technology, Wuhan, China.

[#]These authors contributed equally to this work.

[*]To whom correspondences should be addressed.

E-mails: *sunjiaming@hust.edu.cn; yuanhaowu@hust.edu.cn*

**Abstract**

The hierarchical triple-helix structure of collagen type I (Col I) is essential for extracellular matrix support and integrity. However, current reconstruction strategies face challenges such as chain mismatch, preventing proper fibril formation. Here, we report a supramolecular co-assembly strategy using a *de novo*–designed α-helical peptide amphiphile (APA) of just seven amino acids. The APA features a hydrophobic palmitic acid tail, which stabilizes the helical structure and promotes co-assembly upon interaction with complementary molecular structures. This minimal design enables selective recognition of fragmented collagen (FC), restoring triple-helix conformation and guiding fibril formation. We applied this mechanism to engineer FC-rich nanofat (NF) into a mechanically reinforced biomaterial. Integration of APA-NF with coaxial 3D printing enabled spatial control of structure and function. In a porcine model, this platform enhanced *in situ* vascularized adipose tissue regeneration. Our results demonstrate that hierarchical reconstruction of collagen *via* peptide-guided supramolecular assembly offers a promising strategy for soft tissue repair.

**Introduction**

Extracellular matrix (ECM) regulates fundamental cellular processes[1,2], providing mechanical support[3] and facilitating biological signaling[4–6] by mediating dynamic cell-ECM interactions through mechanotransduction[7,8]. Collagen type I (Col I), the most abundant ECM protein, is essential for maintaining tissue structural integrity[9] and mechanics[10]. These functions are governed by its triple-helix structure[11], composed of three parallel α-polypeptide chains. These chains twist around a central axis to form superhelix molecules, which subsequently organize into fibrils with high axial alignment[12,13]. The critical role of Col I has inspired extensive biomimetic research. The Chaikof group[14] and Xu group[15] have each designed repeated modular collagen-like peptides sequences to synthesize typical D-periodic banded fibrils. Collagen-like trimers as short peptides can replicate the triple-helix motif but have struggled to organize into parallel fibrillar structures[16–19]. Thus, these approaches are limited by the inability to fully reproduce the atypical triple-helix structure of Col I, complicating efforts to recreate its native architecture[20].

The triple-helix structure is stabilized by side-chain hydrogen bonding, which mediates the specific folding and alignment of three α-chains[21]. The resulting Col I monomers undergo hierarchical alignment and bundle into fibrils and subsequently fascicles *via* hydrophobic and electrostatic interactions[12]. This hierarchical fibrillar structure is highly resilient but difficult to repair after mechanical disruption[13,22]. The Yu group has developed collagen hybridizing peptides (CHPs) based on POG repeats, which exhibit strong triple-helical propensity[23], but require thermal pretreatment to prevent self-aggregation[24]. Peptoids reported by the Feng group[25] have been employed to inhibit the CHP self-trimerization and enhance thermal stability, but this intervention has slowed the folding rate[26]. Recently, the Raines group synthesized covalently-tethered parallel CHP strands (dimeric CHPs) to target individual denatured collagen chains[26–28], but these often result in mismatches[29,30]. Moreover, due to their tendency to self-trimerize, their primary role is to target denatured Col I, rather than to mimic native Col I[26]. Consequently, CHP-remodeled Col I lacks the intrinsic driving force necessary for subsequent assembly into fibrils following hybridization[31,32]. Therefore, the challenge in repairing the triple-helix structure of fragmented Col I (FC) lies in ensuring specific non-covalent interactions and achieving hierarchical self-assembly.

Multicomponent self-assembly (co-assembly) enables the programmed, spontaneous organization of monomers into supramolecular architectures through specific non-covalent interactions[33–36]. Our previous work has demonstrated co-assembly yielding distinct structural features such as membranes[37–39] and gel[40]. Extensive research has been conducted on controlling the co-assembly of short peptide amphiphiles (PAs) into well-defined hierarchical structures[40–42] such as fibrils[43–45], ribbons[46,47], and lamellas[48,49]. Most existing co-assembling systems rely on β-sheet motifs to drive ordered molecular alignment[50–52]. In contrast, α-helix-based self-assembly systems are rare, as stabilizing the α-helical conformation demands numerous intra-strand hydrogen bonds. Long peptide sequences (exceeding 21 amino acids[53,54]) offer more hydrogen bonding opportunities but are prone to nonspecific interactions and misfolding. Conversely, shorter sequences (single heptad repeat module) ensure precise molecular interactions but suffer from conformational instability[55]. To address this, conjugating a hydrophobic block to a short peptide may promote and stabilize a well-defined secondary structure intrinsic to the peptide itself[56,57]. Hydrophobic domain interactions drive the self-assembly of PAs into highly ordered structures by inducing aggregation through hydrophobic collapse, thereby increasing the local concentration of conjugated peptide segments. Compared with traditional biosynthetic or mimetic approaches, co-assembly offers a promising strategy for reconstructing the complex hierarchical organization of Col I, spanning from triple-helical nanostructure to aligned fibrillar microstructure.

Nanofat (NF), obtained through mechanical emulsification of adipose tissue, is rich in FC. It contains cells, cytokines, and other bioactive components, granting it inherent advantages such as immune compatibility and broad availability[58]. Owing to its high bioactivity and compatibility, NF has been successfully applied in scar repair, facial rejuvenation, and other regenerative fields[59]. However, NF lacks volumetric support across applications[60] due to the disrupted Col I network within its ECM, compromising its mechanical strength[61]. Researchers have attempted to enhance the mechanical property of NF by physically incorporating materials with greater stiffness, such as ceramic particles[62,63] and hyaluronic acid[64], but leading to displacement and immune rejection due to poor localization. More critically, these materials fail to replicate the dynamic mechanical adaptability of native ECM. Therefore, NF may serves as an ideal autologous biomaterial for evaluating Col I repair strategies[65] and advancing soft tissue regeneration.

Here, we report on an innovative PAs based on short sequence of α-helix peptide with only 7 amino acids (AA) and a hydrophobic palmitic acid (Pal) tail. The unconjugated 7AA peptide adopts a random coil conformation, while conjugation with Pal induced α-helical folding. First, we employed collagenases to develop an enzymatically FC model and validated the targeted repair of Col I mediated by the α-helical peptide amphiphile (APA). Experimental analyses confirmed that the APA could repair Col I fibers by restoring their characteristic triple-helical structure, the mechanism was elucidated by coarse-grained molecular dynamics (CGMD) computer simulations. Second, we engineered NF into a bioactive material with remodeled structure and enhanced mechanical properties through co-assembling with APA, which effectively repaired the disrupted Col I network within NF and augmented its bioactivity. Finally, we integrated the APA-NF system with coaxial 3D printing to fabricate hierarchical bio-scaffolds. *In vivo* studies using nude mice and porcine models demonstrated successful *in situ* remodelling of Col I and regeneration of vascularized adipose tissue. We illustrated the innovative co-assembly mechanism and application of APA-FC/NF system, which demonstrated the possibility to design functional self-assembling PAs based on α-helix.

**Results**

**1.System rationale and materials designing**

APA was designed comprise of a hydrophilic short peptide and a hydrophobic tail (Fig. 1a). Hydrophilic short peptide AE (AEKIRKE) was chosen to repair the FC, which was reported to present a α-helix structure[66,67] within a protein (Fig. 1a). However, the organization of seven residue peptide into a three-dimensional (3D) helical fold was thermodynamically unfavourable[53,68]. To address this, a hydrophobic Pal tail was conjugated to the N-terminal of AE as APA (Pal-AEKIRKE) because Pal has been proved to increase the triple helix folding stability[57,69,70] (Fig. 1a, b). To further validate the triple-helical sequence following the structural insights, we designed PAK (Pal-AEKIREK) and PAR (Pal-AEKIKRE) as controls (Fig. 1b). The reposition of the last three AA disrupted the triple-helix structure, while preserved the total molecular charge and AA composition.

The FC model was established using collagenase to mimic the structural degradation of Col I (Fig. 1c). FC was individually co-assembled with four PAs sequences into nano- and micro-structures to assess their capacity to repair FC toward original Col I. APA was found to successfully co-assemble FC into hydrogel with triple-helix featured structure. Then, NF with an abundance of FC was co-assembled with APA into hydrogel with robust mechanical properties by repairing the Col I network. With the synergistic effect of bioactive components in NF, we rationalized APA-NF (AN) gels would enable adipose regeneration and vascularization (Fig. 1d). Furthermore, we integrated a coaxial nozzle with the APA-NF system to supramolecular biofabricate hierarchical scaffolds for *in vivo* grafting. Finally, nude mice and porcine soft tissue defect models confirmed that the AN gel scaffolds with repaired Col I molecular structure promoted soft tissue regeneration in a physiological microenvironment.

**2.Underlying mechanism of APA-FC co-assembly system**

*Establishing the fragmented FC model*

To elucidate APA guided reconstitution of Col I distinct triple-helical structure in co-assembling processes, we performed collagenolysis to destroy the structure of Col I using metalloproteinase (MMP). Fibrillar Col I was cleaved into characteristic fragments by MMP-1[65,71,72] and further untangled from triple-helical structure into single-strand collagen polypeptides by MMP-9[73] (Fig. 2a). We first demonstrated that the particle size of Col I was significantly reduced from 173.9 ± 16.0 nm into 81.6 ± 24.0 nm after collagenolysis process by Dynamic Light Scattering (DLS). Since MMP-9 unwinds the helical trimer into monomers without changing the length of molecule, there is no significant difference between the size of Col I molecules digested by MMP-1 alone (81.6 ± 24.0 nm) and by both MMP-1&MMP-9 (94.8 ± 13.4 nm). The native Col I displayed a typical triple-helix spectrum of a positive peak around 220 nm and a large negative one near 200 nm[74,75] characterized by circular dichroism (CD, Fig 2b, solid grey line). FC lost the typical triple helix structure after degradation, showed a red shifting of the negative ellipticity to 205 nm with a positive peak at approximately 193 nm, suggesting the presence of untangled helical molecules[76] (Fig. 2b, solid blue line). The FC with random fibrils was revealed by transmission electron microscopy (TEM) and scanning electron microscopy (SEM). The native Col I fibrils were approximately 100 nm in diameter with well-defined transversely banded structure akin to the D-periodicity as reported[15] in TEM images, the length and diameter of FC were significantly reduced through collagenolysis with absence of D-periodicity (Extended Data Fig. 1a). These reductions in diameter and length of Col I fibrils were further confirmed by SEM (Extended Data Fig. 1b). In conclusion, these results demonstrate that we have successfully recreated a model of FC with an unraveled triple-helical structure by collagenolysis using MMP.

*Supramolecular characterization of APA-FC co-assembly*

After recreating the FC, we used APA, AE, PAK, and PAR to co-assemble with FC and try to recreate the triple-helix and D-periodic structure. First, we determined that FC carries a negative charge, which was opposite to that of the APA and control PAs (Fig. 1b) measured by electrophoretic light scattering (ELS). Then, the particle size of APA before and after co-assembly with FC (Fig. 2c) were analysis by DLS. In both collagenolysis stages, FC co-assembled with APA to form particles of approximately 3000 nm, which were significantly larger than those formed with other control PAs, measuring smaller than 1500nm. The formation of the largest co-assembled structures by FC and APA among all tested PAs suggests the strongest intermolecular interactions[40,77] between FC and APA. We studied the secondary structure changes of FC before and after co-assembly using CD. APA exhibited a negative peak at around 202 nm, driving FC to display a blue shift of the negative ellipticity back to 200 nm, providing evidence for the regaining of the original triple-helical structure of Col I (Fig.2b, dashed grey line). By comparison, FC displayed a further red shift with other control PAs (Extended Data Fig. 1c), which deviating from the characterized pattern of Col I. Therefore, these results confirm that APA reconstructs the triple-helix secondary structure of FC through co-assembly.

*Micro- and Nano- scales characterization of APA-FC co-assembling structures*

The co-assembling of APA-FC gave rise to hierarchical architectures spanning from micro- to nano-scale features. FC was able to co-assemble with APA into fibrillar aggregates at microscale in SEM observations with a pronounced increase in fibril diameter (Extended Data Fig. 1b). The features of fibril formation with APA were further visualized by TEM. The FC (MMP-1&9), digested into single chains, co-assembled with APA to form fibrillar bundles characterised by periodic bright and dark bands (Extended Data Fig. 1a). The co-assembled nanostructure features of APA-FC were revealed by small-angle neutron scattering (SANS). By fitting with the flexible cylinder model, APA and FC were characterized as cylindrical nanofibrils with 59-Å radius and 482-Å radius, respectively. After co-assembling, the resulting APA-FC nanostructure was interpreted as a core-shell cylindrical nanofibril with 604-Å radius, with a shell of 75.5-Å in thickness (Fig. 2d and Extended Data Fig. 2a). APA-FC showed pronounced scattering at low-$q$, indicating a cylindrical nanofibril structure. Additionally, APA-FC exhibited scattering at high-$q$ region, which corresponds to aggregates with smooth surfaces, further validating the formation of uniform co-assembled structure (Extended Data Fig. 2b). Overall, these results demonstrate that APA can effectively co-assemble with FC and facilitate the formation of fibrillar bundles with D-periodicity at micro- and nano- scales.

*Intermolecular mechanism of APA-FC co-assembly*

To further elucidate the molecular mechanism underlying the experimentally observed supramolecular co-assembly in APA-FC system, we performed CGMD simulations using the MARTINI forcefield. In designing the coarse-grained representation of FC, we adopted a 29 AA Col I model[78], in which each helical chain is composed of repeating Gly-Pro-Hyp sequences. FC was modelled as monomeric one-quarter and three-quarter fragments of this sequence, reflecting the two-step collagenolysis by MMP-1 and MMP-9 used in experimental processing. Simulations showed that FC alone aggregated into a less well-defined globule (Extended Data Fig. 3). Upon the spatially and rotationally random introduction of APA into the system (Fig. 2e, left), its presence actively promotes the co-assembly of the FC into elongated, fibrillar structures (Fig. 2e, right and Supplementary Movie 1), which are analogous to Col I. These simulation results consistent with the co-assembled morphology observed by SEM (Fig. 2f). To elucidate the molecular basis underlying this structural difference, we examined the structure formation of APA-FC co-assembled unit. Simulations revealed that two FC chains intertwined with the AE segment to form a triple-helical structure, with the hydrophobic Pal core buried at its center (Fig. 2g, right). As part of a multiscale simulation strategy, we performed a smaller-scale all-atom simulation of the APA-FC system to obtain a finer-grained structural representation. Ramachandran plot revealed that the $\Phi$ and $\Psi$ dihedral angles in this atomistic model (Fig. 2g, bottom) closely resemble those characteristic of Col I[79]. Both the coarse-grained and atomistic simulations support the formation of an adaptive supramolecular triple-helical domain within the co-assembled system, with APA functioning as a molecular template that organizes FC chains into triple-helix conformations and promotes long-range order reminiscent of native Col I fibrils.

To further verify the co-assembling mechanism through spatial correlations between APA and FC, we analysed the radial distribution function (RDF) of APA-FC system (Fig.2h). The RDF profiles between FC, AE, Pal, and MARTINI water (W) revealed that FC exhibited strong spatial associations with both the hydrophobic Pal and hydrophilic AE domains of APA, indicating intimate molecular contact. The RDF also indicated that Pal localized at short distance from FC, with the hydrophilic AE and amphiphilic FC components facing the aqueous phase, consistent with hydrophobic core formation. This spatial organization facilitates growth along the fibril axis. Complementarily, the different structural morphology and comparison of structural anisotropy between FC-only and APA-FC systems were shown by the radius of gyration (Rg) time-series analysis (Fig. 2i). The APA-FC system showed much smaller decrease in Rg over time than FC-only system, indicating that more mass remained distributed away from the central axis, suggesting that APA prevents the collapse of FC into globular aggregates and promotes directional co-assembly into collagen-like architectures. Further, both systems exhibited the same aggregation propensity (AP = 2.0)[80,81], indicating that the difference in Rg is not due to a difference in the total degree of aggregation but rather to the shape of the self-/co-assembling structure each system organized into. This molecular mechanism supports the experimentally observed co-assembly behaviour and elucidates the spatial organization of APA and FC that underlies triple-helix formation. This molecular arrangement not only stabilizes the helical core but also facilitates directional co-assembly, ultimately enabling fibril elongation.

## 3. Characterization of the APA-NF co-assembly system

*Supramolecular characterization of APA-NF co-assembly*

After revealing the co-assembling mechanism of APA and FC, APA was utilised to reconnect the FC-rich network within NF. NF carried a negative charge as expected, which was opposite to that of APA identified by ELS (Fig. 3a and Extended Data Fig. 4b). The average molecular size of NF suspension increased significantly from ~252 nm to ~2000 nm after co-assembling with APA evidenced by DLS (Fig. 3b). We further investigated the secondary structure changes associated with APA-NF co-assembly using CD. NF exhibited a typical α-helix spectrum, with a positive peak at 192 nm and negative peaks at 222 nm and 208 nm, transitioned to a triple-helical pattern upon co-assembly with APA (Extended Data Fig. 4c). Inspired by the set-up of supramolecular biofabrication through interfacial liquid-liquid phase separation (LLPS)[37,38,40]. NF was directly injected into the APA water solution to fabricate the APA-NF (AN) gel, where interfacial co-assembly was triggered immediately upon contacting (Fig.1d and Supplementary Movie 2). During this process, the generated AN gel maintained the injection-defined geometry, allowing precise control over its morphology based on injection speed and trajectory. The resulting AN gel was able to be retrieved intact from the APA water solution and maintained its structural integrity under compressive force applied by forceps (Fig. 3c and Supplementary Movie 3). These results confirm APA enables co-assembling with NF into robust hydrogel as a printable system with high fidelity.

*Enhanced mechanical properties of AN gel after co-assembling*

The observed structural integrity of AN gel inspired further investigation into its mechanical properties, which are critical for bioprinting and soft tissue grafting. Accordingly, rheological testing was performed on both AN gel and NF. Both materials were elastic-dominant, with storage moduli (G′) consistently exceeding loss moduli (G″) across 0.1-100 rad s$^{-1}$, shown by frequency sweep measurements. AN gel exhibited markedly higher G′ values than NF at both low (1045.45 Pa vs. 158.91 Pa) and high frequencies (2414.94 Pa vs. 381.52 Pa), suggesting the superior mechanical robustness and stiffness of AN gel under-deformation condition (Fig. 3d, left). Under constant shear (1 s$^{-1}$), AN gel maintained a significantly higher viscosity (207.85 Pa s) than NF (31.06 Pa s) over time, reflecting its superior mechanical integrity and structural cohesion (Extended Data Fig. 5a, top). Both materials showed a crossover point between G′ and G″ near 100% strain under strain sweep from 0.1 to 100% strain at 1 rad s$^{-1}$, indicating comparable yielding behaviour under mechanical loading (Extended Data Fig. 5a, bottom). These findings demonstrate that mechanical reinforcement in AN gel is achieved without sacrificing deformability, ensuring ease of extrusion. Shear-thinning behaviour was observed in both materials, as shown by negative slopes in shear rate sweeps (0.01-100 s$^{-1}$), which is essential for injection-based applications (Fig. 3d, middle). To evaluate self-recovery, a series of low (1%) and high (500%) strains cycles was applied to each material as the G' and G" were measured. Both materials showed modulus inversion under high strain, indicating a transition to liquid-like behaviour, followed by rapid recovery under low strain. AN gel recovered 72% of its original G′ within 1 min, compared to 64% for NF (Fig. 3d, right). This rapid recovery is essential for maintaining fidelity of AN gel structure after printing. These results demonstrate that remodelling FC network within AN gel lead to substantial mechanical reinforcement and improved mechanical recovery, while preserving favourable shear-thinning behaviour essential for printing applications, and yielding a modulus of approximately 1.6 kPa similar to that of native adipose tissue[82,83].

*Micro- and Nano- scale characterization of AN co-assembling gel*

To explore whether the enhanced mechanical properties of AN gel resulted from its hierarchical structure formed during co-assembly, structural and compositional characterizations were carried out at both micro- and nano-scales. First, aligned fibrillar bundles resembling native Col I fibrils were observed on AN gel, whereas NF showed disordered aggregates, as visualized by SEM (Fig. 3e). This indicates that the insertion process facilitates directional remodelling of

FC within NF, which is critical for reconstructing a native ECM-like architecture. In order to assess the components and key cytokines within AN gel, the total protein concentration was measured using the bicinchoninic acid (BCA) assay. As expected, the protein concentration in AN gel was similar to that of NF, indicating the co-assembly process preserved the protein components of NF in AN gel (Fig. 3f). AN gel and NF exhibited significantly higher concentration than fat as the emulsion protocol condensed the protein components[84]. Then, the components of NF and AN gel were analysed quantitatively by sodium dodecyl sulfate-polyacrylamide gel electrophoresis (SDS-PAGE) and Western blot (WB). AN gel presented all bands representing in APA and NF, revealing that all components within APA and NF were incorporated in AN gel (Fig. 3g). This approach also illustrated that APA successfully co-assembled with NF at very low concentrations (0.5%), and full co-assembling reaction completed in just 5 minutes (Extended Data Fig. 5b). Notably, several certain key cytokines known positively regulate cell proliferation and vascularization[85–89] appeared as darker bands in AN gel compared to NF in the WB analysis, suggesting that the co-assembling procedure condense the NF phase through interfacial diffusion-reaction[90] (Fig. 3h). In conclusion, co-assembly of APA and NF rapidly rebuild the Col I fibrillar network and condense with bioactive components, providing both structural and biochemical potential for tissue regenerative applications.

**4. Biological validation**

*Aggregation of cellular components via co-assembly*

The successful co-assembly of AN gel with fibrillar matrix and endogenous active components highlighted its potential as a bioactive material for tissue regeneration. In addition to abundant cytokines, NF contains a high density of vascular stromal cells, notably adipose-derived stem cells (ADSCs)[91]. To investigate cellular organization during the co-assembly process, we first performed nuclear staining to visualize cell distribution within NF and AN gel. Fluorescence microscopy revealed that the cells were significantly aggregated into clusters within the AN gel, whereas sparsely dispersed in the NF (Fig. 4a), suggesting that the co-assembling process promotes localized cellular congregation. This spatially favourable aggregation of cells inspired further evaluation of the biocompatibility and biological effects of AN gel.

*Biocompatibility of AN gel*

The biocompatibility of AN gel was first assessed the potential toxicity. The absence of genotoxicity was confirmed by comet assay results, with the percentage of DNA in the comet tail for ADSCs co-cultured with NF or AN gel remaining below 5% (Fig. 4b), comparable to the negative control (untreated cells). Subsequently, we validated the biocompatibility and proliferative effects of AN gel by co-culturing with ADSCs and human umbilical vein endothelial cells (hUVECs) separately for 7 days, given their distinct roles in adipose tissue regeneration and angiogenesis[92,93]. The experiments were performed in parallel with cells co-cultured with NF and an untreated control group (Extended Data Fig. 6a). The biocompatibility of AN gel was further supported by Live/Dead assays. Both ADSCs and hUVECs maintained morphology characteristic of metabolically active cells over culture period, with few dead cells observed (Extended Data Fig. 6b). Continuous proliferation in both ADSCs and hUVECs across all groups were revealed by cytotoxicity assays. Notably, ADSCs exhibited significantly higher viability in the AN gel group at day 3 compared to both NF and control. At day 7, no statistical difference was observed between the AN gel and the NF groups for either cell type (Fig. 4c). These results demonstrate AN gel is non-genotoxic and can serve as a biocompatible material.

*Supporting adipogenesis and angiogenic sprouting in vitro*

The demonstrated biocompatibility of AN gel motivated further validation of cellular responsiveness to its bioactive components. Therefore, we further validated the effect of AN gel in inducing functional cellular activity. Adipogenic lineages differentiation in ADSCs is a key process in adipose tissue formation. The formation and fusion of lipid droplets during 21-day adipogenic induction in ADSCs co-cultured with AN gel, NF, or blank control group were evaluated by BODIPY staining (Fig. 4d). Quantitatively analysis revealed that the AN gel group exhibited the highest accumulation of lipid droplets in both number (4356 cells per field) and area ($21.38 \times 10^4$ μm$^2$), compared to the NF group (2501 cells per field; $9.35 \times 10^4$ μm$^2$) and blank control group (743 cells per field; $3.71 \times 10^4$ μm$^2$) (Fig. 4e). Furthermore, the expression of adipogenic genes involved in lipid metabolism (*GLUT4, Perilipin1, LPL, CEBPa, and PPARγ*)[94] were markedly upregulated in the AN gel group relative to both NF and blank control, supporting the enhanced adipogenic potential of AN gel (Fig. 4f).

To verify the ability of AN gel to enhance cell migration, the key to embryonic blood vessel formation[95–98], hUVECs were used to assess the response preference towards active components in AN gel. HUVECs were seeded on a porous insert membrane with immersed in leaching solution of AN gel, NF, or untreated medium. The number of migrated hUVECs on the bottom of the membrane were independently measured after 24 hours. AN gel was the most effective group in promoting cells migration (610 cells per filed), compared to the NF group (510 cells per filed) and blank control group (417 cells per filed) (Fig. 4g). After 3 days of co-culture with AN, NF, or blank control group, hUVECs co-cultured with AN gel exhibited significantly higher expression of genes (*PLAUR, CXCR4, PFKFB, and APLN*), which were associated with sprouting behaviour of hUVECs[99] (Fig. 4h). Taken together, the results demonstrate that the AN gel has the potential to direct cells behaviour related to soft tissue regeneration and angiogenesis *in vitro*.

## 5. Multi-scale supramolecular biofabrication *via* 3D printing

*Coaxial 3D printing enables micro-to-macro control*

We developed the techniques to guide multiscale co-assembly process with APA-NF system using covalent extrusion 3D printer, according to the co-assembly method and the previous rheological findings. NF was extruded through the inner core, continuously co-assembling with the APA water solution from the outer core during printing (Fig. 5a). A series of 3D structures with increasing complexity was printed with APA-NF system. A 4-layers 5×5 log pile that designed to have a length and width of 25 mm was successfully printed, which had 10 pores unobstructed and each layer had been precisely deposited. For the stable G′ after co-assembling, the collapse of structure and the deformation of the bottom layer was not observed. To increase complexity, G-code was manually written to print the letters "WHUH" and modified Hilbert Curve, which had continuous turning and folding in 5 mm, each curve was successfully printed with each turning was clearly visible (Fig. 5b and Supplementary Movie 4). The co-assembling process reconstructs aligned fibrils when coaxially printed under the horizontal direction guidance. The relation between the edge of AN gel scaffolds and printing direction were observed by SEM. Images show that the fibrillar bundles were well organized in alignment with the long axis of the printed direction (Fig. 5c). These results demonstrate that the APA-NF system achieves controlled multi-scale supramolecular biofabrication with coaxial 3D printer, enabling simultaneous precision in aligned fibrillar microstructure reconstruction and complex 3D macrostructural fidelity.

*AN gel scaffold promoting vascular morphogenesis*

To evaluate the capacity of the AN gel scaffold to support cell adhesion and 3D growth through its realigned microarchitecture. We seeded ADSCs (membrane-labeled, red) and hUVECs (autofluorescent, GFP) in equal proportions ($10^7$ mL$^{-1}$) on AN gel scaffold to observe cell-cell and cell-scaffold dynamic interactions. Confocal images showed initial cell adhesion at 8 hours after seeding, with cells maintaining a round morphology and beginning to extend pseudopodia after 24 hours. Both ADSCs and hUVECs exhibited intercellular contact by day 3, particularly hUVECs, forming tube-like structures within AN gel. The number of viable cells increased with time, leading to the formation of a continuous tubular vascular-like structure by day 7 (Fig. 5d). The morphology of the cell cluster on the final cell-scaffold construct was visualized using immunostaining of cell surface markers. Images reinforced that hUVECs arranged into a vessel-mimicking lumen, shaped by the topological features of AN gel scaffold (Fig. 5e). These results demonstrate that, by taking advantage of supramolecular force driven Col I-like fibrils reconstruction and enhanced bioactivity of the APA-NF system, make it feasible to fabricate 3D scaffolds that mimic native ECM in multiscale and promote cell growth.

## 6. Evaluation of soft tissue regeneration *in vivo*

*Structural and morphological evaluation of grafted scaffolds*

We established a subcutaneous graft model in nude mice to evaluate the *in vivo* regeneration capacity of AN gel. To assess both the short-term degradation profile and the long-term regenerative performance of AN gel scaffolds, we performed histological and morphological analyses at 4 and 12 weeks post-grafting. The *in vivo* status of the grafted scaffolds was assessed at both time points, revealing regenerative adipose tissue with subdermal vascular connection in the AN gel group, whereas the NF group exhibited structural fragmentation and increased oil cysts formation containing liquefied fat (Extended Data Fig. 7b). At both 4 and 12 weeks post-grafting, the AN gel group exhibited significantly greater volume retention (57.73% and 42.20%, respectively) compared to the NF group (46.63% and 22.20%) (Extended Data Fig. 7c). The AN gel group preserved structural integrity over time, maintaining adipose lobule architecture and minimizing vacuole formation, with portions of the 3D-printed spiral shape still visible at 4 weeks in Hematoxylin and Eosin (H&E) staining images. In contrast, the NF group showed structural fragmentation and acellular cavities (Fig. 6a and Extended Data Fig. 8). The dense and organized Col I was deposited in the AN gel group, while the NF group exhibited progressive disintegration shown by Masson's Trichrome (Masson) and Immunohistochemistry (IHC) staining. In the AN gel group, collagen content increased from 14.45% pre-grafting to 50.66% at 12 weeks post-grafting, while the NF group peaked at 14.07% at 4 weeks and declined to 12.85% at 12 weeks (Fig. 6b). Quantification of Col I-positive areas showed a similar trend at 4 and 12 weeks post-grafting, increased from 14.28% to 45.63% in the AN gel group, but declined from 12.87% to 12.16% in the NF group, despite similar baseline levels before grafting (Fig. 6b). These findings were further confirmed with aligned fibrillar structures and intact, spherical cells observed in AN gel scaffolds by SEM at 12 weeks (Fig. 6c, purple arrows), whereas the NF group showed disorganized fibrils and collapsed cells with irregular membranes (Fig. 6c, yellow arrow). These results validate the capability of AN gel to maintain morphology after grafting *in vivo* while providing a favourable tissue microenvironment by reconstructing ECM that promotes tissue regeneration, especially for cells and Col I.

*Adipocyte maturation and vascularization of grafted scaffolds in vivo*

To demonstrate that the cells grown in the scaffolds were indeed adipocytes and their correlation with vascularization during the repair process, immunofluorescence (IF) staining for Perilipin (PLIN, adipocyte marker) and CD31 (endothelial marker) were performed (Fig. 6a and Extended Data Fig. 8,9). PLIN-positive areas in the AN gel group increased from 1.26% pre-grafting to 1.65% at 4 weeks post-grafting and 6.98% at 12 weeks post-grafting, while the NF group decreased from 1.25% to 0.04% and slightly increased to 0.84%. Consistently, CD31-positive areas in the AN gel group rose from 0.51% to 0.95% and 1.31% at 4 and 12 weeks post-grafting, whereas the NF group declined from 0.52% to 0.16% and remained low at 0.18%. These results indicated that adipocytes progressively matured with a concomitant increase in adipocyte density over the 12 weeks post-grafting (Fig. 6b). We also observed blood vessels formed by endothelial cells surrounding vacuolated adipocytes (Extended Data Fig. 8,9), establishing pathways for nutrient diffusion for adipose tissue regeneration. These findings confirm that AN gel promotes sustained adipocyte development in parallel with vascular formation after grafting *in vivo*.

*Preclinical porcine soft tissue defect model to demonstrate the repair function of AN gel*

To further evaluate the efficacy of the AN gel in clinical application, we conducted a preclinical pilot study using a porcine model. This model aimed to address physiological and anatomical differences between small animals and humans, particularly related to size and inflammatory response. We developed a soft tissue defect porcine model by surgically removing 3 mm × 3 mm squares of 1 cm-thick subcutaneous adipose tissue from the lateral abdomen. The injured area underwent fibrotic repair leading to scar tissue healing after one-month of surgical wound stabilization. Conventional autologous filler materials macrofat (diameter >1 mm) and NF as control groups, because they were widely used in clinic. They were compared to the AN gel, and each filler was performed on the site of the soft tissue defects. The grafted site and adjacent intact soft tissue were collected at 6 months post-surgery, to characterize the architecture of dermis and subdermal adipose tissue. Concurrently, untreated defective site tissues were harvested as controls to characterize scar-mediated healing.

Acute loss of subdermal adipose tissue triggered fibrotic remodelling, characterized *in vivo* by well-demarcated depression and dense fibrous deposition. While all treatment groups exhibited restoration in varying degrees, only the AN gel group demonstrated near-complete volumetric recovery with seamless tissue integration and indistinct boundaries between regenerated and native regions (Fig. 6d, blue overlays). In contrast, macrofat and NF treatments resulted in residual depressions, suggestive of limited structural remodelling. AN gel grafting fully restored tissue volume and supported the regeneration of large, mature adipocyte clusters (Extended Data Fig. 10, blue arrows) throughout the subdermal layer, closely recapitulating native morphology, as shown by H&E staining. Importantly, the fibrotic tissue (Fig. 6d, green overlays) was replaced by a well-organized structure, as clearly revealed by Masson staining specific to fibrous tissue, suggesting that AN gel disrupted scar-dermis adhesions and facilitated hierarchical tissue remodelling during regeneration. Although partial adipose regeneration was observed in the macrofat and NF group, both failed to resolve persistent scar-dermis adhesions or re-establish native tissue architecture. The histological findings were further corroborated by evidence of effective scar remodelling in the AN gel group, as demonstrated by IHC staining for Col I (Fig. 6d). Together, these results demonstrate that AN gel promotes structural regeneration by replacing disorganized scar tissue with organized, layered regenerative tissue, achieving both volumetric restoration and functional integration beyond passive space-filling repair.

*Adipocyte morphology and vascular distribution differences in tissue regeneration*

We further looked into the *in situ* generation at cell level, regenerated adipocytes in the AN gel group (Fig. 6d, yellow arrows) exhibited clear, continuous PLIN-positive membranes and distributed in lobular clusters in IF staining of PLIN (green), which were resembled with those in native adipose tissue (Fig. 6d, white arrows). In contrast, other groups showed sparse PLIN staining, with atrophic adipocytes and disrupted spatial continuity, indicating incomplete adipose tissue regeneration. Tissue vascularization and significant differences in vascular distribution were further characterized. The AN gel group exhibited uniform vascular distribution resembling native tissue, characterized by IF staining of CD31 (Fig. 6d, red). Notably, a high-density vascular network (Fig. 6d, yellow asterisks) was observed at the dermal-fat interface, suggesting active angiogenic support for regenerating adipocytes. In control, macrofat, and NF groups, scar tissue regions contained densely clustered small-diameter blood vessels, while sparse vascularization surrounding adipocyte clusters. These results demonstrate that AN gel actively promotes coordinated regeneration of adipose tissue and vasculature, effectively recapitulates native organization compared to other treatments.

**Discussion**

In this study, we demonstrate the rational design of a functional APA capable of reconstructing the hierarchical triple-helix structure of Col I through supramolecular co-assembly with FC. Despite comprising only seven amino acids, APA exhibits a strong intrinsic α-helical folding propensity. The conjugated hydrophobic palmitic acid tail plays a pivotal role in stabilizing this conformation and promoting supramolecular assembly. This molecular design enabled the remodelling of FC, significantly enhancing both the biological functionality and mechanical properties of NF, as validated *in vitro* and *in vivo*. Furthermore, the integration of APA-NF system with coaxial 3D printing established a programmable supramolecular biofabrication platform. Key advantages of this work include: (1) *de novo* design of α-helical PAs; (2) hierarchical reconstruction *via* co-assembly; (3) functional translation into biomaterials; and (4) a clinically translatable fabrication strategy. Collectively, these findings represent a significant advance in supramolecular design, leveraging native self-assembly mechanisms to engineer functional biomaterials for regenerative medicine.

**Methods**

**Synthesis and characterization of PAs**

APA, AE, PAK, and PAR molecules were provided by GL biochem (China). The sequences and molecular weights of the PAs are shown in Figure 1a. PAs were synthesized via solid-phase chemical synthesis and characterized by reversed-phase high-performance liquid chromatography and electrospray ionization mass spectrometry.

**Nuclear magnetic resonance spectroscopy (NMR)**

NMR spectroscopy was used to investigate the solution-state conformation of the APA. All spectra were acquired at 298K on a Bruker Avance III 600 MHz superconducting NMR spectrometer (Bruker, Germany) equipped with a triple-resonance cryogenic probe. Samples were prepared in a buffer consisting of 90% $H_2O$ and 10% $D_2O$ at a final APA concentration of 4 mM. For resonance assignments and conformational analysis, a standard suite of two-dimensional (2D) NMR experiments was performed, including $^1H$-$^1H$ Correlation Spectroscopy (COSY), $^1H$-$^1H$ Total Correlation Spectroscopy (TOCSY), and 1H-1H Nuclear Overhauser Effect Spectroscopy (NOESY), using standard Bruker pulse programs. Specifically, the experiments were a gradient-selected, phase-sensitive COSY (cosygpmfppqf), a TOCSY utilizing a DIPSI-2 spin-lock (dipsi2esgpph) with a mixing time of 150 ms, and a NOESY (noesyesgpph) with a mixing time of 300 ms. NMR data were processed using TopSpin 3.6.2 and NMRPipe software[100]. The processed NMR spectra were visualized, assigned and analysed using POKY software[101].

**Collagenolysis of Col I**

Purified rat tail collagen type I sterile solution (5 mg/mL in 20 mM acetic acid) was purchased from Bio-Techne, US. Fibril formation was induced by diluting collagen into 1X PBS and adjusting the pH to neutral with 1N NaOH. Final collagen concentration was 0.4 mg/mL, and the solution was incubated at 37 °C for 1 h. MMP1 and MMP9 (ACRO Biosystems, China) were activated using 1 mM p-aminophenylmercuric acetate (APMA, Sigma, US) in TCNB buffer (Sigma, US) at 50 μg/mL and 100 μg/mL, respectively, at 37 °C. Activated enzymes were diluted to 0.2 μg/mL (MMP1) and 1 μg/mL (MMP9) in TCNB buffer and incubated with Col I at 37 °C overnight.

**Dynamic light scattering (DLS)**

Particle size distributions of Col I and FC were measured using a Nanolink instrument (linkoptik, China). Samples were diluted to 0.05 wt% with distilled water prior to analysis.

**Circular dichroism (CD)**

Solutions of Col I and FC (0.05 wt%) were transferred into 0.01 mm cuvettes and analysed using a CD spectrometer (JASCO, Japan). Spectra were collected by averaging 10 scans over the 190-260 nm range at a scan rate of 50 nm/min. Data were smoothed using a simple moving average method.

**Transmission electron microscopy (TEM)**

Col I and FC solutions were deposited onto 200-mesh carbon-coated TEM grids for 5 min, followed by staining with 2 wt% uranyl acetate for 30 s. Grids were rinsed with ultrapure water for 30 s and air-dried at room temperature for 24 h. Bright-field imaging was performed on an HT7800 TEM (HITACHI, Japan) operating at 80 kV.

**Scanning electron microscopy (SEM)**

Samples of Col I and FC were fixed with 4% paraformaldehyde (PFA, Servicebio, China) for 10 min and dehydrated using an ethanol gradient (70%, 80%, 90%, 96%, 100%) at room temperature. Dried samples were coated with gold after critical point drying (HARVENT, China) and imaged using GeminiSEM 300 (ZEISS, Germany).

**Electrophoretic light scattering (ELS)**

Zeta potentials of Col I (0.05 wt%) and PAs (0.01%) solutions were measured at 25 °C using a NanoLink zeta potential analyzer (linkoptik, China). Samples were prepared in ultrapure water and equilibrated for 10 min at the measurement temperature.

**Small-angle neutron scattering (SANS)**

$D_2O$ was added to an equal volume of FC solution. The PAs were dissolved in $D_2O$ at 2 wt%, and the FC (50% $D_2O$/50% $H_2O$)-PAs (2 wt%) mixed solution was prepared for measurement. Measurements were performed using the SANS instrument (The BL01 at China Spallation Neutron Source, China). Solutions (1 mL) of individual components were loaded into 1-mm-path-length hellma quartz cuvettes, while the mixture was prepared in a demountable 1-mm-path-length cuvette. Cuvettes were mounted in aluminum holders within a sealed, computer-controlled sample chamber maintained at 25 °C. Each measurement took approximately 30 min. All scattering data were normalized to sample transmission, background correction was performed using a quartz cell filled with $D_2O$, and detector linearity and efficiency were corrected using the instrument-specific software. The incident neutrons with wavelength of 1.1-9.8 Å were defined by a double-disc bandwidth chopper, which is collimated to the sample by a pair of apertures. The experiment used the sample to detector distance of 5 m and a sample aperture of 8 mm in diameter. The two dimensional $^3$He tubes array detector allows to cover a wide Q range from 0.004 Å$^{-1}$ to 0.7 Å$^{-1}$. We collected approximate 60 min scattering information for each sample, including the empty sample holder and sample cell. The scattering data were set to absolute unit after normalization, transmission correction and standard sample calibration.

**Molecular Dynamics Simulations**

Coarse-grained molecular dynamics (CGMD) simulations were performed using the MARTINI forcefield (version 2.2)[102–104] via the GROMACS 2025.2 software package[105,106]. Systems consisted of randomly distributed FC peptides, with or without APA, in a pre-equilibrated MARTINI water box. Simulations were run under standard MARTINI protocols with standard temperature, pressure, and electrostatics settings. The structural evolution of FC alone and APA+FC mixtures was monitored to evaluate their respective assembly behaviours. Parameters for aggregation propensity (AP), radius of gyration (Rg), and radial distribution functions (RDFs) were extracted from the trajectories. Full simulation details are provided in the Supplementary Methods.

In simulating the all-atom system, the CHARMM36 forcefield[107,108] was employed, utilising a series of simulation steps including minimization, NVT, NPT with Berendsen pressure coupling[109], and production with Parrinello-Rahman pressure coupling. The LINCS algorithm[110,111] was applied to constrain all hydrogen atom bonds, enabling an increase in the timestep from 1 fs to 2 fs. The entire simulation was conducted over a total duration of 100 ns. To expedite the interactions between APA and FC, a weak flat-bottom restraint was introduced between one of the APA monomers and the other monomers. Particle Mesh Ewald[112] was used to evaluate electrostatic interactions throughout the atomistic simulation.

**Sample preparation (NF)**

Adipose tissue was obtained from abdominoplasty patients with informed consent and ethics approval (Huazhong University of Science and Technology, [2019] IEC (S864)). After mincing, samples were centrifuged at 2000 × g for 3 min. NF was prepared by emulsifying the tissue between two 10-mL syringes connected with Luer Lock connectors (Tulip, US) of decreasing diameters (1.2 mm, 1.0 mm, 0.8 mm).

**Sample preparation (APA-NF gel)**

APA was dissolved in deionized water to a final concentration of 2 wt%. An equal volume of NF solution was slowly injected into the APA solution at a constant rate without stirring. The mixture was then left undisturbed at 4 °C for 1 hour to allow complete gel formation. The resulting APA-NF gel was collected for subsequent use.

**Bicinchoninic acid (BCA) assay**

Fat tissue was minced and homogenized to obtain fat homogenates. For fat, NF, and AN gel samples, RIPA lysis buffer (Thermo Fisher Scientific, US) with PMSF (Aladdin, China) was added at 250 μL per 20 mg sample. After lysis on ice, lysates were centrifuged at 14,000 × g for 5 min, and supernatants collected. Protein levels were quantified using the BCA Protein Assay Kit (Thermo Fisher Scientific, US). Diluted BSA standards and samples were added to a 96-well plate with BCA working reagent (sample to working reagent ratio = 1:8), incubated at 37 °C for 30 min, and absorbance measured at 540-590 nm. Protein concentrations were calculated based on the standard curve.

**Sodium dodecyl sulfate-polyacrylamide gel electrophoresis (SDS-PAGE)**

For SDS-PAGE, AN gel was prepared under two experimental designs: (1) fixed reaction time of 1 h with APA at concentrations of 2.0%, 1.5%, 1.0%, 0.5%, 0.1%, 0.05%, and 0.01%; (2) fixed APA concentration of 1% with reaction times of 5, 10, 30, and 60 min. After gelation, residual APA was blotted off. Gels were disrupted by pipetting and vortexing, then diluted 10-fold in PBS. Samples (15 μL of fat, NF, AN, and APA solution) were mixed with 5 μL of loading buffer (Sigma, US), heated at 90 °C for 5 min, and loaded into pre-cast NuPAGE Bis-Tris gels (ACE Biotechnology, China). A 5 μL pre-stained protein ladder (10-180 kDa, Thermo Fisher Scientific, US) was included. Electrophoresis was conducted for 1 h at 120 V using NuPAGE MES SDS buffer, followed by Coomassie brilliant blue staining (Beyotime, China) and imaging with a gel documentation system (BLT-IMAGING, China).

### Western blot (WB)

Proteins from fat, NF, and AN gel were separated by SDS–PAGE and transferred to PVDF membranes. Membranes were blocked in 5% nonfat milk in TBST for 30 min at room temperature and incubated overnight at 4 °C with primary antibodies (Servicebio, China) diluted in 3% BSA in TBST. After washing, membranes were incubated with secondary antibodies (Servicebio, China) for 1 h at room temperature and developed using an enhanced chemiluminescence system.

### Rheological test

Rheological measurements were performed on a Discovery Hybrid Rheometer (TA Instruments, US) using a 20 mm parallel plate. A total of 500 μL of NF or AN gel was deposited onto the stage, and the top plate was lowered to a 1000 μm gap. Excess material was removed. The tests included: frequency sweep (0.1-100 rad $s^{-1}$ at 1% strain), time sweep (at a constant shear rate of 1 $s^{-1}$), strain sweep (0.1-100% strain at 1 rad $s^{-1}$), shear rate sweep (0.01-100 $s^{-1}$), and shear recovery (a minute cycles each at high (500%), low (1%) strain at 1 rad $s^{-1}$).

### Isolation and culture of ADSCs

Complete ADSC culture medium consisted of L-DMEM (Gibco, US), 10% FBS (HyClone, US), and 1% penicillin/streptomycin (PS, Gibco, US). Adipose tissue was washed with sterile PBS and digested with 0.2% collagenase NB-4 (Nordmark, Germany) in complete medium at 37 °C for 4 h. After digestion, the tissue was centrifuged at 1500 rpm for 5 min. The cell suspension was filtered and seeded into culture dishes. Non-adherent cells were removed on day 3 by medium exchange. Cells at 80-90% confluence were passaged using trypsin/EDTA (Thermo Fisher Scientific, US). ADSCs from passages 3-5 were used for all experiments.

### DAPI staining

To visualize the spatial distribution of cells in NF and AN gel, samples were fixed in 4% PFA for 20 min at room temperature, followed by three PBS washes. Fixed samples were stained with DAPI (Beyotime, China) and incubated at room temperature before imaging by confocal microscopy (Nikon, Japan).

**Comet assay**

ADSCs were co-cultured with either NF or AN gel for 1 week. Untreated cells were used as negative controls; cells treated with 25 μM potassium permanganate for 20 min served as positive controls. Cells were dissociated into single-cell suspensions, adjusted to $1 \times 10^5$ cells/mL, mixed with 1% low-gelling-temperature agarose at 37 °C (1:6 v/v), and loaded onto agarose-coated slides. Slides were immersed in lysis buffer (Bio-Techne, US) for 1 h at 4 °C, then washed with PBS for 30 min twice at room temperature. Slides were placed in rinse buffer and subjected to electrophoresis at 20 V (0.6 V/cm) for 30 min. After neutralization in distilled water, slides were stained with 1.0 μg/mL ethidium bromide (Thermo Fisher Scientific, US) for 20 min. All images were performed using confocal microscopy and analysed with ImageJ.

**Cell viability and cytotoxicity testing**

hUVECs were obtained from Procell (China) and cultured in RPMI medium (Gibco, US) supplemented with 10% FBS (Gibco, US) and 1% PS (Gibco, US). ADSCs or hUVECs were seeded on porous insert membranes with test materials placed on top and cultured for 7 days. Cell viability was assessed using Live/Dead Cell Double Staining Kit (Sigma, US) on days 1, 3, and 7. A mixture of propidium iodide (PI) and Calcein AM in PBS was applied for 30 min at 37 °C. Cells were visualized using a confocal microscope. Cytotoxicity was evaluated using a Cell Counting Kit 8 (Abcam, UK) following the manufacturer's protocol, and absorbance was measured at 450 nm using a microplate reader (BioTek, US).

**Adipogenic induction of ADSCs**

For adipogenic differentiation, ADSCs were treated with adipogenic induction medium (AIM), consisting of H-DMEM (Gibco, US), 10% FBS (HyClone, US), 1% PS (Gibco, US), 1 μM dexamethasone (Sigma, US), 200 μM indomethacin (Sigma, US), 10 μg/mL insulin (Sigma, US), 2 μM rosiglitazone (Sigma, US) and 0.5 mM IBMX (Sigma, US). To maintain differentiation, adipogenic maintenance medium (AMM) was composed of H-DMEM, 10% FBS, 1% PS, 2% L-glutamine, and 10 μg/mL insulin. On day 7 of co-culture, adipogenic induction was initiated by alternating AIM (3 days) and AMM (1 day).

**BODIPY and phalloidin staining:**

Cells were washed three times with PBS and fixed with 4% PFA for 30 min at room temperature. They were then incubated with 2 μM BODIPY (MedChemExpress, US) in the dark for 30 min, followed by three PBS washes. Alexa Fluor 594-phalloidin (Thermo Fisher Scientific, US) diluted in PBS was added and incubated in the dark for 60 min. Nuclei were counterstained with DAPI for 20 min. Samples were washed with PBS and imaged using a confocal microscope (Nikon, Japan).

## PCR

ADSCs were co-cultured with NF or AN gel for 7 days for adipogenesis analysis, and hUVECs were co-cultured for 1 day to assess sprouting. Total RNA was extracted using TriQuick reagent (Solarbio, China) and stored at −80 °C. Reverse transcription was performed using 2.5 μg of RNA with a SCILOGEX SCI1000-G system. Quantitative PCR was conducted on an ABI QuantStudio 1 real-time PCR system (Thermo Fisher Scientific, US) using primers listed in Table S1. PCR cycling conditions were: 94 °C for 2 min, followed by 45 cycles of 94 °C for 5 s and 60 °C for 30 s. GAPDH was used as the housekeeping gene. Relative expression levels were determined using the ΔΔCt method and normalized to the control group.

## Chemotactic migration assay

Chemotactic migration of hUVECs was assessed using a Transwell assay (8.0 μm pore size, Corning, US). HUVECs were seeded at a density of $5 \times 10^3$ cells per insert in serum-free medium and incubated for 24 hours at 37°C. The lower chambers contained leachate (experimental) or basal medium (control). After 24 hours, non-migrated cells on the upper surface of the membrane were gently removed with a cotton swab. The membrane was then removed from the insert, and DAPI staining was performed to visualize migrated cells on the underside.

## APA-NF system coaxial printing

A coaxial nozzle was used to extrude the APA-NF system, with NF as the core fluid (17G needle) and APA solution as the sheath fluid (13G needle). Printing paths were controlled *via* G-code and executed using BioMaker V2 software on a Bioprinter (SunP Biotech, China). Printing speed and extrusion rate were set to 5 mm/s and 5 mm³/s, respectively. APA solution was driven by a plunger at 50 mm/min.

## Cell seeding on scaffold

For live-cell tracking, GFP-hUVECs (IMMOCELL, China) were used. ADSCs were labeled with the membrane dye DiI (red fluorescence) and pseudocolored in magenta during image processing to distinguish from live/dead staining. AN gel scaffolds were printed into sterile 6-well plates and pre-wetted with medium for 2 h. ADSCs and hUVECs were mixed 1:1 (final concentration: $1 \times 10^7$ cells/mL) in a 1:1 blend of their complete media. Cell suspension was pipetted onto scaffolds and incubated under standard conditions. Cell distribution was visualized by confocal microscopy (Nikon, Japan).

## Immunofluorescence staining

Scaffolds co-cultured with cells for 7 days were fixed with 4% PFA and processed for paraffin embedding. Sections were deparaffinized, rehydrated, and blocked with 3% BSA for 30 min at room temperature. Primary antibodies (Servicebio, China) against CD29 (for ADSCs) and CD31 (for hUVECs) were applied overnight at 4 °C. After washing, secondary fluorophore-conjugated antibodies (Service, China) were added for 50 min at room temperature. DAPI was used for nuclear staining. Samples were visualized using a confocal microscope.

**Animal experiments (nude mice model)**

Animal procedures were approved by the Institutional Animal Care and Use Committee of Huazhong University of Science and Technology ([2023] IACUC Number:4735). Nude mice were anesthetized with 30 mg/kg sodium pentobarbital (Sigma, US). To evaluate this, spiral layer-by-layer 3D printed AN gel scaffold was subcutaneously grafted over the dorsum of the nude mice. The control group consisted of NF scaffolds formed by replacing the outer-axis solution with sterile saline solution. Each scaffold had a fixed volume of 0.25 mL, with standardized printing parameters minimizing inter-sample variability. Prior to grafting, the dorsal skin of nude mice was sterilized. Two small incisions (approximately 10 mm) were made on either side of the dorsal midline, followed by blunt dissection of the subcutaneous tissue beneath each incision to create independent subcutaneous pockets. Care was taken to ensure that the two pockets did not communicate, thereby preventing scaffold displacement or fusion. AN gel and NF scaffolds were then inserted into the respective subcutaneous cavities above the muscle. The incisions were closed with sutures. After surgery, the mice were placed in an incubator until full recovery from anesthesia. Tissues were harvested at 4 and 12 weeks post-grafting (n = 4 per group) following euthanasia. Histological assessment was performed by blinded observers.

**Histological staining**

Subcutaneously grafts were harvested and fixed in 4% PFA overnight at 4 °C. The tissue specimen was sectioned in the cross-sectional or longitudinal direction and stained with haematoxylin and eosin (Servicebio, China) and Masson's trichrome (Servicebio, China). For Masson-stained images, collagen deposition was evaluated by calculating the collagen area (%) as the ratio of blue-stained area to the total stained area in randomly selected fields. Immunohistochemical staining for Col I was performed, and the positive area of Col I was quantified based on the stained regions. Immunofluorescence staining was conducted as described previously, targeting Perilipin (PLIN) and CD31. The percentage of PLIN-postive cells and CD31-positive cells was quantified in randomly selected regions. All image analysis was performed using ImageJ.

**Animal experiments (porcine model)**

Animal procedures were approved by the Committee on Animal Care of Hubei Yizhicheng Biotechnology Co., Ltd. (Number WDRM-202401003). Female pigs weighing 30-35 kg were anesthetized by inhalation of isoflurane, followed by endotracheal intubation to maintain anesthesia. Throughout the procedure, respiratory rate, heart rate, blood pressure, and blood oxygen saturation were continuously monitored. To establish soft tissue defects, animals were placed in a lateral recumbent position and secured. The lateral abdominal skin was disinfected thoroughly. A rectangular skin flap measuring 12 cm in length and 8 cm in width was created on the abdominal region, with the long edge serving as the pedicle. The flap was dissected at the level below the deep adipose layer to allow full mobilization. Four square defects (3 mm × 3 mm) were created within the flap, each dissected down to a depth of 1 cm beneath the dermis. All defects were spatially separated to prevent intercommunication and ensure independent wound healing. The incisions were closed and the flap secured using the tie-over bolster technique to provide compression and prevent the formation of subcutaneous dead space. Postoperatively, animals received antibiotic prophylaxis for 3 days to prevent infection.

One month after the initial surgery, the incision sites had healed well, and visible depressions were observed at the defect locations, with evident adhesion to the underlying tissues. Following the same preoperative preparation and anesthesia protocol as before, pigs were placed in the prone position and the surgical site on the posterior neck was sterilized. An incision was made along the longitudinal axis of the neck, and an adequate volume of subcutaneous fat tissue was harvested. The incision was closed and compressed using a pressure dressing. The harvested fat was cut with surgical scissors into injectable macrofat for subsequent use, and remaining portion further processed into NF and AN gel. After releasing adhesions between the dermis and underlying tissues at each defect site, the three types of fillers were injected from the edge of each defect using a syringe. A volume of 7 mL was injected per defect, slightly overfilling the depression to a level just above the skin surface. The injection sites were closed with sutures. Postoperatively, animals received antibiotic prophylaxis for 3 days to prevent infection.

**Statistical analysis**

Statistical analysis was performed using one-way or two-way ANOVA followed by Tukey's post-hoc test using GraphPad Software (Prism v.10). P values were used to determine significant differences (NS at $P > 0.05$, $*P < 0.05$, $**P < 0.01$, $***P < 0.001$, and $****P < 0.0001$).

**References**

1. Chaudhuri, O., Cooper-White, J., Janmey, P. A., Mooney, D. J. & Shenoy, V. B. Effects of extracellular matrix viscoelasticity on cellular behaviour. *Nature* **584**, 535–546 (2020).

2. Vogel, V. & Sheetz, M. Local force and geometry sensing regulate cell functions. *Nat Rev Mol Cell Biol* **7**, 265–275 (2006).

3. Humphrey, J. D., Dufresne, E. R. & Schwartz, M. A. Mechanotransduction and extracellular matrix homeostasis. *Nat Rev Mol Cell Biol* **15**, 802–812 (2014).

4. Kechagia, J. Z., Ivaska, J. & Roca-Cusachs, P. Integrins as biomechanical sensors of the microenvironment. *Nat Rev Mol Cell Biol* **20**, 457–473 (2019).

5. Panciera, T., Azzolin, L., Cordenonsi, M. & Piccolo, S. Mechanobiology of YAP and TAZ in physiology and disease. *Nat Rev Mol Cell Biol* **18**, 758–770 (2017).

6. Oria, R. *et al.* Force loading explains spatial sensing of ligands by cells. *Nature* **552**, 219–224 (2017).

7. Saraswathibhatla, A., Indana, D. & Chaudhuri, O. Cell – extracellular matrix mechanotransduction in 3D. *Nat Rev Mol Cell Biol* **24**, 495–516 (2023).

8. Discher, D. E., Janmey, P. & Wang, Y. Tissue Cells Feel and Respond to the Stiffness of Their Substrate. *Science* **310**, 1139–1143 (2005).

9. Sorushanova, A. *et al.* The Collagen Suprafamily: From Biosynthesis to Advanced Biomaterial Development. *Advanced Materials* **31**, 1801651 (2019).

10. Mouw, J. K., Ou, G. & Weaver, V. M. Extracellular matrix assembly: a multiscale deconstruction. *Nat Rev Mol Cell Biol* **15**, 771–785 (2014).

11. Ricard-Blum, S. The Collagen Family. *Cold Spring Harbor Perspectives in Biology* **3**, a004978–a004978 (2011).

12. Gelse, K. Collagens — structure, function, and biosynthesis. *Advanced Drug Delivery Reviews* **55**, 1531–1546 (2003).

13. Shoulders, M. D. & Raines, R. T. Collagen Structure and Stability. *Annu. Rev. Biochem.* **78**, 929–958 (2009).

14. Rele, S. *et al.* D-Periodic Collagen-Mimetic Microfibers. *J. Am. Chem. Soc.* **129**, 14780–14787 (2007).

15. Hu, J. *et al.* Design of synthetic collagens that assemble into supramolecular banded fibers as a functional biomaterial testbed. *Nat Commun* **13**, 6761 (2022).

16. Frank, S. *et al.* Stabilization of short collagen-like triple helices by protein engineering. *Journal of Molecular Biology* **308**, 1081–1089 (2001).

17. Gauba, V. & Hartgerink, J. D. Self-Assembled Heterotrimeric Collagen Triple Helices Directed through Electrostatic Interactions. *J. Am. Chem. Soc.* **129**, 2683–2690 (2007).

18. Gauba, V. & Hartgerink, J. D. Surprisingly High Stability of Collagen ABC Heterotrimer: Evaluation of Side Chain Charge Pairs. *J. Am. Chem. Soc.* **129**, 15034–15041 (2007).

19. Gauba, V. & Hartgerink, J. D. Synthetic Collagen Heterotrimers: Structural Mimics of Wild-Type and Mutant Collagen Type I. *J. Am. Chem. Soc.* **130**, 7509–7515 (2008).

20. O'Leary, L. E. R., Fallas, J. A., Bakota, E. L., Kang, M. K. & Hartgerink, J. D. Multi-hierarchical self-assembly of a collagen mimetic peptide from triple helix to nanofibre and hydrogel. *Nature Chem* **3**, 821–828 (2011).

21. Yammine, K. M. *et al.* An outcome-defining role for the triple-helical domain in regulating collagen-I assembly. *Proc. Natl. Acad. Sci. U.S.A.* **121**, e2412948121 (2024).

22. Zhang, Y., Herling, M. & Chenoweth, D. M. General Solution for Stabilizing Triple Helical Collagen. *J. Am. Chem. Soc.* **138**, 9751–9754 (2016).

23. Persikov, A. V., Ramshaw, J. A. M., Kirkpatrick, A. & Brodsky, B. Amino Acid Propensities for the Collagen Triple-Helix. *Biochemistry* **39**, 14960–14967 (2000).

24. Wang, A. Y., Mo, X., Chen, C. S. & Yu, S. M. Facile Modification of Collagen Directed by Collagen Mimetic Peptides. *J. Am. Chem. Soc.* **127**, 4130–4131 (2005).

25. Goodman, M., Melacini, G. & Feng, Y. Collagen-Like Triple Helices Incorporating Peptoid Residues. *J. Am. Chem. Soc.* **118**, 10928–10929 (1996).

26. Ratnatilaka Na Bhuket, P., Li, Y. & Yu, S. M. From Collagen Mimetics to Collagen Hybridization and Back. *Acc. Chem. Res.* **57**, 1649–1657 (2024).

27. Tanrikulu, I. C., Westler, W. M., Ellison, A. J., Markley, J. L. & Raines, R. T. Templated Collagen "Double Helices" Maintain Their Structure. *J. Am. Chem. Soc.* **142**, 1137−1141 (2020).

28. Li, X., Zhang, Q., Yu, S. M. & Li, Y. The Chemistry and Biology of Collagen Hybridization. *J. Am. Chem. Soc.* **145**, 10901–10916 (2023).

29. Fiala, T. *et al.* Frame Shifts Affect the Stability of Collagen Triple Helices. *J. Am. Chem. Soc.* **144**, 18642–18649 (2022).

30. Li, Y. & Yu, S. M. Targeting and mimicking collagens via triple helical peptide assembly. *Current Opinion in Chemical Biology* **17**, 968–975 (2013).

31. Strauss, K. & Chmielewski, J. Advances in the design and higher-order assembly of collagen mimetic peptides for regenerative medicine. *Current Opinion in Biotechnology* **46**, 34–41 (2017).

32. He, L. & Theato, P. Collagen and collagen mimetic peptide conjugates in polymer science. *European Polymer Journal* **49**, 2986–2997 (2013).

33. Finkelstein-Zuta, G. *et al.* A self-healing multispectral transparent adhesive peptide glass. *Nature* **630**, 368–374 (2024).

34. Huang Wu *et al.* Dynamic supramolecular snub cubes. *Nature* **637**, 347–353 (2025).

35. Bianco, S. *et al.* Mechanical release of homogenous proteins from supramolecular gels. *Nature* **631**, 544–548 (2024).

36. Thanapongpibul, C. *et al.* Unlocking Intracellular Protein Delivery by Harnessing Polymersomes Synthesized at Microliter Volumes using Photo-PISA. *Advanced Materials* **36**, 2408000 (2024).

37. Wu, Y. *et al.* Disordered protein-graphene oxide co-assembly and supramolecular biofabrication of functional fluidic devices. *Nat Commun* **11**, 1182 (2020).

38. Wu, Y., Fortunato, G. M. & Okesola, B. O. An interfacial self-assembling bioink for the manufacturing of capillary-like structures with tuneable and anisotropic permeability. *Biofabrication* **13**, 035027 (2021).

39. Wu, Y. *et al.* Disinfector-Assisted Low Temperature Reduced Graphene Oxide-Protein Surgical Dressing for the Postoperative Photothermal Treatment of Melanoma. *Adv Funct Materials* **32**, 2205802 (2022).

40. Wu, Y. *et al.* Co-assembling living material as an in vitro lung epithelial infection model. *Matter* **7**, 216–236 (2024).

41. Clemons, T. D. & Stupp, S. I. Design of materials with supramolecular polymers. *Progress in Polymer Science* **111**, 101310 (2020).

42. Kaygisiz, K., Sementa, D., Athiyarath, V., Chen, X. & Ulijn, R. V. Context dependence in assembly code for supramolecular peptide materials and systems. *Nat Rev Mater* (2025) doi:10.1038/s41578-025-00782-6.

43. Pashuck, E. T., Cui, H. & Stupp, S. I. Tuning Supramolecular Rigidity of Peptide Fibers through Molecular Structure. *J. Am. Chem. Soc.* **132**, 6041–6046 (2010).

44. Mondal, S., Rehak, P., Ghosh, N., Král, P. & Gazit, E. Linear One-Dimensional Assembly of Metal Nanostructures onto an Asymmetric Peptide Nanofiber with High Persistence Length. *ACS Nano* **16**, 18307–18314 (2022).

45. Jain, A. *et al.* Connected Peptide Modules Enable Controlled Co-Existence of Self-Assembled Fibers Inside Liquid Condensates. *J. Am. Chem. Soc.* **144**, 15002–15007 (2022).

46. Deechongkit, S., Powers, E. T., You, S.-L. & Kelly, J. W. Controlling the Morphology of Cross β-Sheet Assemblies by Rational Design. *J. Am. Chem. Soc.* **127**, 8562–8570 (2005).

47. Cui, H., Cheetham, A. G., Pashuck, E. T. & Stupp, S. I. Amino Acid Sequence in Constitutionally Isomeric Tetrapeptide Amphiphiles Dictates Architecture of One-Dimensional Nanostructures. *J. Am. Chem. Soc.* **136**, 12461–12468 (2014).

48. Zhang, S. *et al.* A self-assembly pathway to aligned monodomain gels. *Nature Mater* **9**,

594–601 (2010).

49. Okesola, B. O. *et al. De Novo* Design of Functional Coassembling Organic–Inorganic Hydrogels for Hierarchical Mineralization and Neovascularization. *ACS Nano* **15**, 11202–11217 (2021).

50. Bera, S., Mondal, S., Rencus-Lazar, S. & Gazit, E. Organization of Amino Acids into Layered Supramolecular Secondary Structures. *Acc. Chem. Res.* **51**, 2187–2197 (2018).

51. Versluis, F., Marsden, H. R. & Kros, A. Power struggles in peptide-amphiphile nanostructures. *Chem. Soc. Rev.* **39**, 3434 (2010).

52. Clarke, D. E., Pashuck, E. T., Bertazzo, S., Weaver, J. V. M. & Stevens, M. M. Self-Healing, Self-Assembled β-Sheet Peptide–Poly(γ-glutamic acid) Hybrid Hydrogels. *J. Am. Chem. Soc.* **139**, 7250–7255 (2017).

53. Mondal, S. *et al.* Formation of functional super-helical assemblies by constrained single heptad repeat. *Nat Commun* **6**, 8615 (2015).

54. Dong, H., Paramonov, S. E. & Hartgerink, J. D. Self-Assembly of α-Helical Coiled Coil Nanofibers. *J. Am. Chem. Soc.* **130**, 13691–13695 (2008).

55. Amit, M., Yuran, S., Gazit, E., Reches, M. & Ashkenasy, N. Tailor‐Made Functional Peptide Self‐Assembling Nanostructures. *Advanced Materials* **30**, 1707083 (2018).

56. Yu, Y.-C., Tirrell, M. & Fields, G. B. Minimal Lipidation Stabilizes Protein-Like Molecular Architecture. *J. Am. Chem. Soc.* **120**, 9979–9987 (1998).

57. Egli, J., Esposito, C., Müri, M., Riniker, S. & Wennemers, H. Influence of Lipidation on the Folding and Stability of Collagen Triple Helices—An Experimental and Theoretical Study. *J. Am. Chem. Soc.* **143**, 5937–5942 (2021).

58. Zheng, H. *et al.* Fat extract improves fat graft survival via proangiogenic, anti-apoptotic and pro-proliferative activities. *Stem Cell Res Ther* **10**, 174 (2019).

59. Van Nieuwenhove, I. *et al.* Soft tissue fillers for adipose tissue regeneration: From hydrogel development toward clinical applications. *Acta Biomaterialia* **63**, 37–49 (2017).

60. La Padula, S. *et al.* Nanofat in Plastic Reconstructive, Regenerative, and Aesthetic Surgery: A Review of Advancements in Face-Focused Applications. *JCM* **12**, 4351 (2023).

61. Uyulmaz, S., Sanchez Macedo, N., Rezaeian, F., Giovanoli, P. & Lindenblatt, N. Nanofat Grafting for Scar Treatment and Skin Quality Improvement. *Aesthetic Surgery Journal* **38**, 421–428 (2018).

62. Han, Z. *et al.* Nanofat functionalized injectable super-lubricating microfluidic microspheres for treatment of osteoarthritis. *Biomaterials* **285**, 121545 (2022).

63. Huang, R.-L. *et al.* Dispersion of ceramic granules within human fractionated adipose tissue to enhance endochondral bone formation. *Acta Biomaterialia* **102**, 458–467 (2020).

64. Lv, X. *et al.* Comparative Efficacy of Autologous Stromal Vascular Fraction and Autologous Adipose-Derived Mesenchymal Stem Cells Combined With Hyaluronic Acid for the Treatment of Sheep Osteoarthritis. *Cell Transplant* **27**, 1111–1125 (2018).

65. Adhikari, A. S., Chai, J. & Dunn, A. R. Mechanical Load Induces a 100-Fold Increase in the Rate of Collagen Proteolysis by MMP-1. *J. Am. Chem. Soc.* **133**, 1686–1689 (2011).

66. Ganar, K. A. *et al.* Phase separation and ageing of glycine-rich protein from tick adhesive. *Nat. Chem.* **17**, 186–197 (2024) doi:10.1038/s41557-024-01686-8.

67. Liu, Y., Gilchrist, A. E. & Heilshorn, S. C. Engineered Protein Hydrogels as Biomimetic Cellular Scaffolds. *Advanced Materials* 2407794 (2024) doi:10.1002/adma.202407794.

68. Pomerantz, W. C. *et al.* Lyotropic Liquid Crystals Formed from ACHC-Rich β-Peptides. *J. Am. Chem. Soc.* **133**, 13604–13613 (2011).

69. Egli, J., Siebler, C., Köhler, M., Zenobi, R. & Wennemers, H. Hydrophobic Moieties Bestow Fast-Folding and Hyperstability on Collagen Triple Helices. *J. Am. Chem. Soc.* **141**, 5607–5611 (2019).

70. McGuinness, K., Khan, I. J. & Nanda, V. Morphological Diversity and Polymorphism of Self-Assembling Collagen Peptides Controlled by Length of Hydrophobic Domains. *ACS Nano* **8**, 12514–12523 (2014).

71. Bertini, I. *et al.* Structural Basis for Matrix Metalloproteinase 1-Catalyzed Collagenolysis. *J. Am. Chem. Soc.* **134**, 2100–2110 (2012).

72. Xia, W. *et al.* Expression of catalytically active matrix metalloproteinase‐1 in dermal fibroblasts induces collagen fragmentation and functional alterations that resemble aged human skin. *Aging Cell* **12**, 661–671 (2013).

73. Vanderstichele, S. & Vranckx, J. J. Anti-fibrotic effect of adipose-derived stem cells on fibrotic scars. *WJSC* **14**, 200–213 (2022).

74. Luo, J. & Tong, Y. W. Self-Assembly of Collagen-Mimetic Peptide Amphiphiles into Biofunctional Nanofiber. *ACS Nano* **5**, 7739–7747 (2011).

75. Lopes, J. L. S., Miles, A. J., Whitmore, L. & Wallace, B. A. Distinct circular dichroism spectroscopic signatures of polyproline II and unordered secondary structures: Applications in secondary structure analyses. *Protein Science* **23**, 1765–1772 (2014).

76. Greenfield, N. J. Using circular dichroism spectra to estimate protein secondary structure. *Nat Protoc* **1**, 2876–2890 (2006).

77. Inostroza-Brito, K. E. *et al.* Co-assembly, spatiotemporal control and morphogenesis of a

hybrid protein–peptide system. *Nature Chem* **7**, 897–904 (2015).

78. Gautieri, A., Russo, A., Vesentini, S., Redaelli, A. & Buehler, M. J. Coarse-Grained Model of Collagen Molecules Using an Extended MARTINI Force Field. *J. Chem. Theory Comput.* **6**, 1210–1218 (2010).

79. Traub, A. Y. W. Polymers of Tripeptides as Collagen Models. *J.Mol.Biol.* **43**, 461-477 (1969)

80. Van Teijlingen, A. & Tuttle, T. Beyond Tripeptides Two-Step Active Machine Learning for Very Large Data sets. *J. Chem. Theory Comput.* **17**, 3221–3232 (2021).

81. Frederix, P. W. J. M., Ulijn, R. V., Hunt, N. T. & Tuttle, T. Virtual Screening for Dipeptide Aggregation: Toward Predictive Tools for Peptide Self-Assembly. *J. Phys. Chem. Lett.* **2**, 2380–2384 (2011).

82. Guimarães, C. F., Gasperini, L., Marques, A. P. & Reis, R. L. The stiffness of living tissues and its implications for tissue engineering. *Nat Rev Mater* **5**, 351–370 (2020).

83. Alkhouli, N. *et al.* The mechanical properties of human adipose tissues and their relationships to the structure and composition of the extracellular matrix. *American Journal of Physiology-Endocrinology and Metabolism* **305**, E1427–E1435 (2013).

84. Quintero Sierra, L. A. *et al.* Highly Pluripotent Adipose-Derived Stem Cell–Enriched Nanofat: A Novel Translational System in Stem Cell Therapy. *Cell Transplant* **32**, 09636897231175968 (2023).

85. Rey, S. & Semenza, G. L. Hypoxia-inducible factor-1-dependent mechanisms of vascularization and vascular remodelling. *Cardiovascular Research* **86**, 236–242 (2010).

86. Heldin, C.-H. & Westermark, B. Mechanism of Action and In Vivo Role of Platelet-Derived Growth Factor. *Physiological Reviews* **79**, 1283–1316 (1999).

87. Pugh, C. W. & Ratcliffe, P. J. Regulation of angiogenesis by hypoxia: role of the HIF system. *Nat Med* **9**, 677–684 (2003).

88. Pepper, M. S. Transforming growth factor-beta: Vasculogenesis, angiogenesis, and vessel wall integrity. *Cytokine & Growth Factor Reviews* **8**, 21–43 (1997).

89. Olsson, A.-K., Dimberg, A., Kreuger, J. & Claesson-Welsh, L. VEGF receptor signalling ? in control of vascular function. *Nat Rev Mol Cell Biol* **7**, 359–371 (2006).

90. Bramham, J. E. & Golovanov, A. P. Temporal and spatial characterisation of protein liquid-liquid phase separation using NMR spectroscopy. *Nat Commun* **13**, 1767 (2022).

91. Sesé, B., Sanmartín, J. M., Ortega, B., Matas-Palau, A. & Llull, R. Nanofat Cell Aggregates: A Nearly Constitutive Stromal Cell Inoculum for Regenerative Site-Specific Therapies.

*Plastic & Reconstructive Surgery* **144**, 1079–1088 (2019).

92. Piard, C. *et al.* 3D printed HUVECs/MSCs cocultures impact cellular interactions and angiogenesis depending on cell-cell distance. *Biomaterials* **222**, 119423 (2019).

93. Qin, Y. *et al.* An Update on Adipose‐Derived Stem Cells for Regenerative Medicine: Where Challenge Meets Opportunity. *Advanced Science* **10**, 2207334 (2023).

94. Fang, Q. *et al.* Adipocyte-derived stem cell-based gene therapy upon adipogenic differentiation on microcarriers attenuates type 1 diabetes in mice. *Stem Cell Res Ther* **10**, 36 (2019).

95. Michaelis, U. R. Mechanisms of endothelial cell migration. *Cell. Mol. Life Sci.* **71**, 4131–4148 (2014).

96. Jin, S. *et al.* Conditioned medium derived from FGF-2-modified GMSCs enhances migration and angiogenesis of human umbilical vein endothelial cells. *Stem Cell Res Ther* **11**, 68 (2020).

97. Barkefors, I. *et al.* Endothelial Cell Migration in Stable Gradients of Vascular Endothelial Growth Factor A and Fibroblast Growth Factor 2. *Journal of Biological Chemistry* **283**, 13905–13912 (2008).

98. Gerhardt, H. *et al.* VEGF guides angiogenic sprouting utilizing endothelial tip cell filopodia. *The Journal of Cell Biology* **161**, 1163–1177 (2003).

99. Graney, P. L. *et al.* Macrophages of diverse phenotypes drive vascularization of engineered tissues. *Sci. Adv.* **6**, eaay6391 (2020).

100. Delaglio, F. *et al.* NMRPipe: A multidimensional spectral processing system based on UNIX pipes. *J Biomol NMR* **6**, (1995).

101. Lee, W., Rahimi, M., Lee, Y. & Chiu, A. POKY: a software suite for multidimensional NMR and 3D structure calculation of biomolecules. *Bioinformatics* **37**, 3041–3042 (2021).

102. Marrink, S. J., Risselada, H. J., Yefimov, S., Tieleman, D. P. & De Vries, A. H. The MARTINI Force Field: Coarse Grained Model for Biomolecular Simulations. *J. Phys. Chem. B* **111**, 7812–7824 (2007).

103. Monticelli, L. *et al.* The MARTINI Coarse-Grained Force Field: Extension to Proteins. *J. Chem. Theory Comput.* **4**, 819–834 (2008).

104. De Jong, D. H., Baoukina, S., Ingólfsson, H. I. & Marrink, S. J. Martini straight: Boosting performance using a shorter cutoff and GPUs. *Computer Physics Communications* **199**, 1–7 (2016).

105. Abraham, M. J. *et al.* GROMACS: High performance molecular simulations through

multi-level parallelism from laptops to supercomputers. *SoftwareX* **1–2**, 19–25 (2015).

106. Van Der Spoel, D. *et al.* GROMACS: Fast, flexible, and free. *J Comput Chem* **26**, 1701–1718 (2005).

107. MacKerell, A. D. *et al.* All-Atom Empirical Potential for Molecular Modeling and Dynamics Studies of Proteins. *J. Phys. Chem. B* **102**, 3586–3616 (1998).

108. Bjelkmar, P., Larsson, P., Cuendet, M. A., Hess, B. & Lindahl, E. Implementation of the CHARMM Force Field in GROMACS: Analysis of Protein Stability Effects from Correction Maps, Virtual Interaction Sites, and Water Models. *J. Chem. Theory Comput.* **6**, 459–466 (2010).

109. Berendsen, H. J. C., Postma, J. P. M., Van Gunsteren, W. F., DiNola, A. & Haak, J. R. Molecular dynamics with coupling to an external bath. *The Journal of Chemical Physics* **81**, 3684–3690 (1984).

110. Hess, B., Bekker, H., Berendsen, H. J. C. & Fraaije, J. G. E. M. LINCS: A linear constraint solver for molecular simulations. *J. Comput. Chem.* **18**, 1463–1472 (1997).

111. Hess, B. P-LINCS: A Parallel Linear Constraint Solver for Molecular Simulation. *J. Chem. Theory Comput.* **4**, 116–122 (2008).

112. Darden, T., York, D. & Pedersen, L. Particle mesh Ewald: An $N \cdot \log(N)$ method for Ewald sums in large systems. *The Journal of Chemical Physics* **98**, 10089–10092 (1993).

**Acknowledgements**

This work was supported by the National Natural Science Foundation of China (NSFC) for the Excellent Young Scientists Fund (Overseas, 0214530013), NSFC for Distinguished Young Scholars (82302837), the China Aerospace Science and Technology Corporation (0231530004), the Strategic Partnership Research Funding (HUST-Queen Mary University of London, No.2022-HUST-QMUL-SPRF-07). We thank the Medical Subcentre of Huazhong University of Science and Technology (HUST) Analytical & Testing Centre. This project was approved by the China Spallation Neutron Source (CSNS) under the grant number P0123122900034 and we thank the staff members of the Small Angle Neutron Scattering (https://cstr.cn/31113.02.CSNS.SANS) at the CSNS (https://cstr.cn/31113.02.CSNS), for providing technical support and assistance in data collection.

**Author contributions**

Y.W., J.S., S.S. and J.Y. conceived the project. S.S. and J.Y. carried out the experiments. Y.W. and J.S. supervised the study. G.Z. performed biological characterization. Z.Y. performed the

AN gel co-assembling characterization. Y.C. performed animal experiments. A.v.T. conducted computer simulations. D.Y. conducted the SANS and analysed the data. T.L. conducted the NMR and analysed the data. Y.K. and H.Y. assisted with SANS. H.Z. assisted with NMR. J.C helped perform the animal experiments.

**Competing interests**

The authors declare no competing interests.

**Correspondence and requests for materials**

should be addressed to Yuanhao Wu and Jiaming Sun.

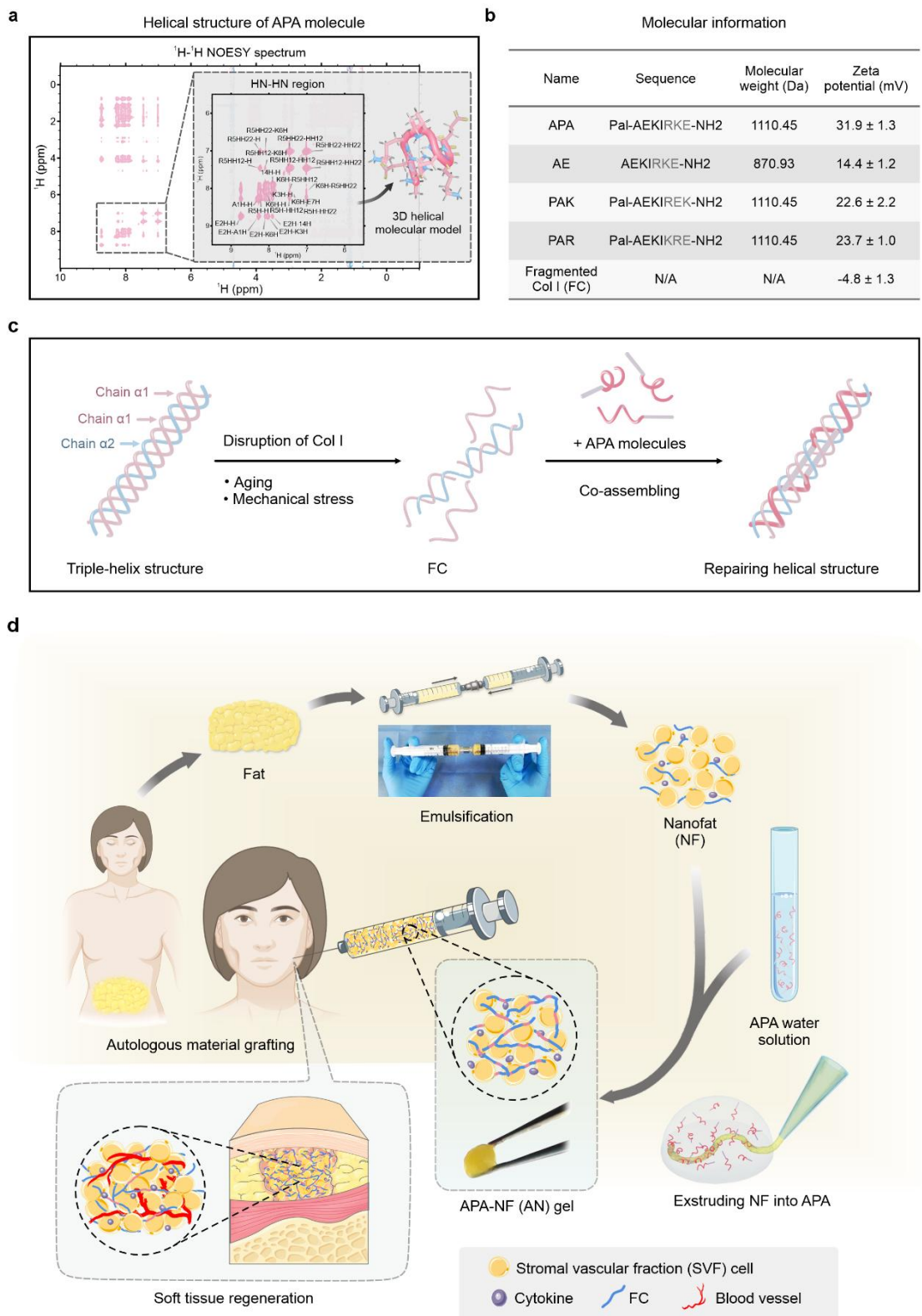

**Fig. 1 | Rationale for co-assembling system and its supramolecular biofabrication applications. a**, $^1$H-$^1$H NOESY spectrum (left) and a representative 3D molecular model (right) of the α-helical peptide (APA). The inset displays an expansion of the amide proton (HN-HN) region. The observation of key medium-range NOEs, such as between residues i and i+3/i+4,

is characteristic of a helical conformation forming in solution. **b**, Table summarizing the key information of the four peptide amphiphiles (PAs) molecules and fragmented Col I (FC) used in the study. **c**, Illustrations of the triple-helix structure disruption of collagen type 1 (Col I) into FC, and structural repairing of FC after co-assembling with APA. **d**, Flowchart of the steps used to prepare nanofat (NF) and fabricate the AN gel as auto-biomaterial. Parts of elements in panel **d** was created using BioRender (https://biorender.com).

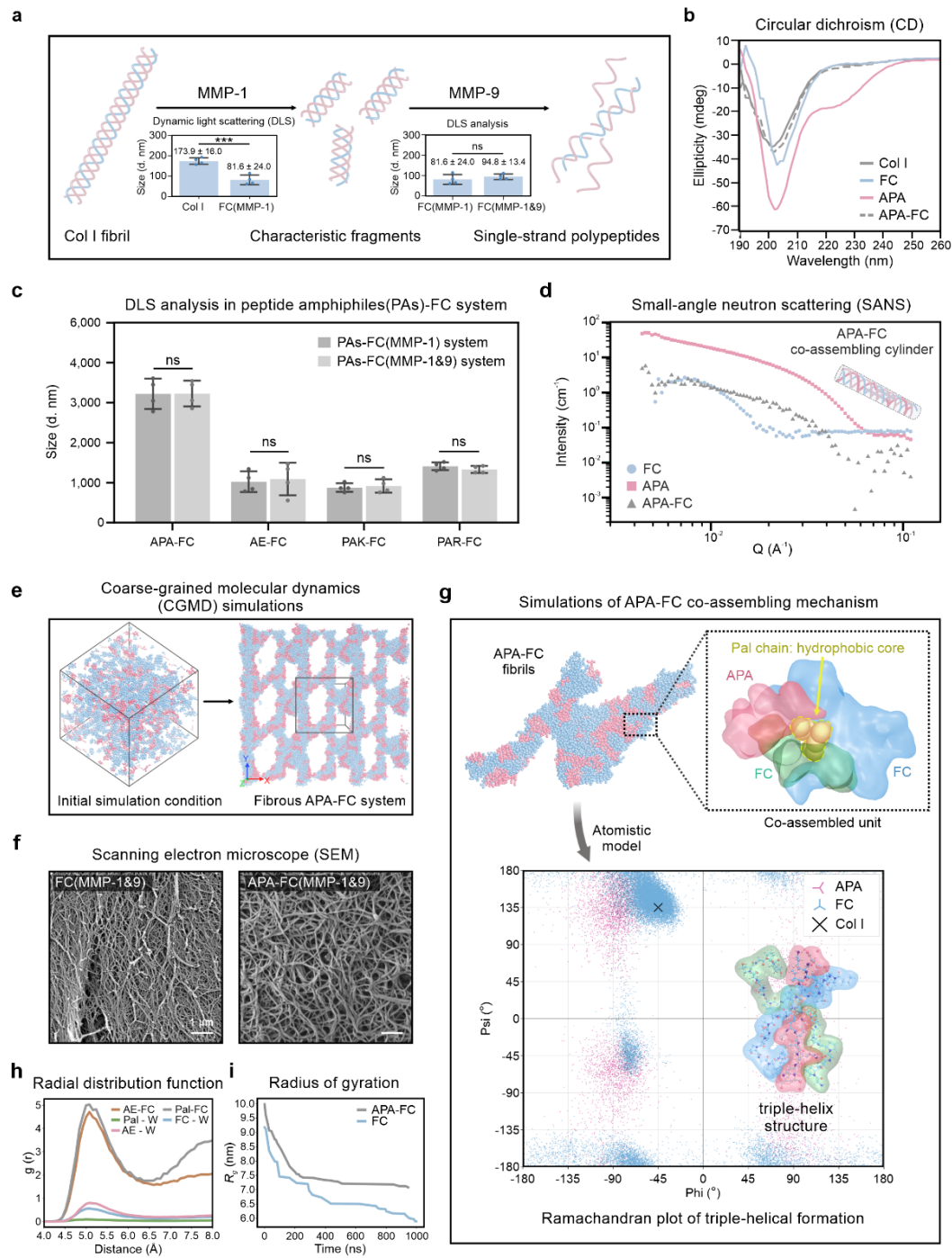

**Fig. 2 | Establishing the fragmented FC model and supramolecular assembly of APA-FC system. a**, Schematic of stepwise Col I degradation into single-stranded FC by collagenolysis with metalloproteinase (MMP). Dynamic light scattering (DLS) analysis showing intermediate size changes. Data are presented as mean ± s.d. (n = 4) **b**, Circular dichroism (CD) spectra demonstrating secondary structure features of Col I and APA-FC system between 190-260 nm. **c**, DLS analysis revealing the intermediate sizes of co-assemblies formed by FC (MMP-1) and FC (MMP-1&9) with each of the four PAs molecules. Data are presented as mean ± s.d. (n = 4). **d**, Small-angle neutron scattering (SANS) patterns of 482-Å radius FC (blue circle) and 59-

Å radius APA (pink square) cylindrical nanofibrils, and a resulting uniform core-shell cylindrical nanofiber of APA-FC co-assembly (grey triangle). **e,** Initial coarse-grained molecular dynamics (CGMD) simulation setup with FC (blue) and APA (pink) randomly distributed in water (water not shown, left), and resulting fibrillar APA-FC structure across periodic boundaries. **f**, Scanning electron microscopy (SEM) images displaying the microstructure of FC (MMP-1&9) before and after co-assembly with APA. **g**, Simulation structure of APA (pink)-FC (cyan) co-assembled unit with the Pal chain (yellow overlays) forming the triple-helix core. Ramachandran plot (bottom) indicates collagenic nature, insert: the atomistic triple helix structure. **h**, Radial distribution function of the APA-FC components illustrating co-assembly between the individual Pal, FC, AE and water (W). **i**, Radius of gyration over time showing the elongated APA-FC compared to the globular FC-only structures. NS at $P >0.05$; ***$P<0.001$ by one-way AVONA and Tukey's post-hoc test.

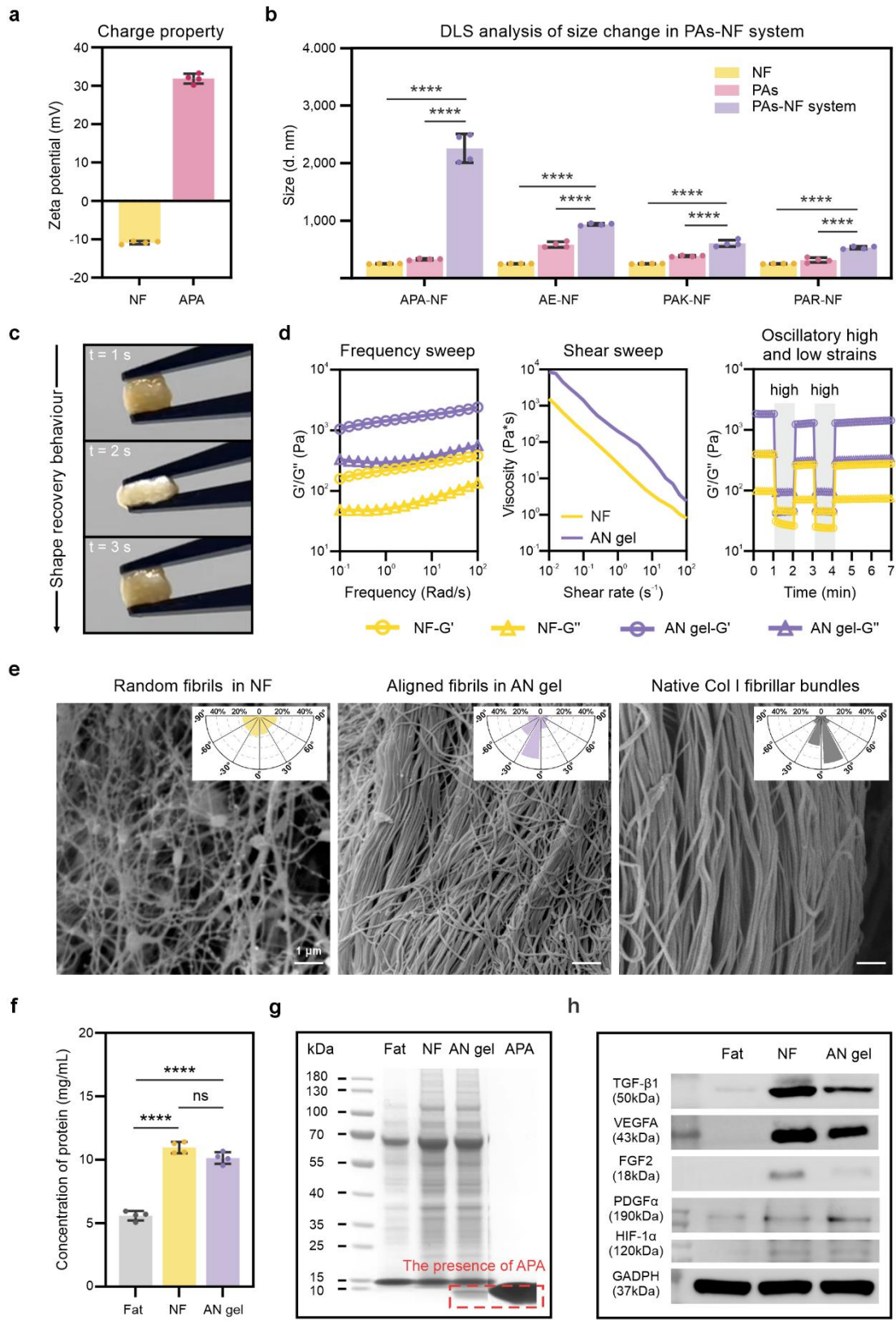

**Fig. 3 | Properties of the co-assembled AN gel. a**, Zeta potential analysis of NF and APA. Data are presented as mean ± s.d. (n = 4). **b**, DLS anslysis revealing the intermediate size change of NF suspension before and after co-assembly with APA, compared to the control PAs. Data

are presented as mean ± s.d. (n = 4). **c**, Recovery of AN gel after mechanical pinching. **d**, Shear sweeps from 0.01 to 100 s$^{-1}$ (left), frequency sweeps from 0.1 to 100 rad s$^{-1}$ (middle) and oscillatory high and low strains (right, white regions represent 1% strain and grey regions represent 500% strain.) of NF and AN. **e**, SEM images of disorganized fibrillar network with large interfibrillar spaces in NF (left), and reconstruced into co-assembled fibrils within AN gel (middle), which is resemble to native Col I (right). **f**, Quantification of total protein content in fat, NF and AN gel using bicinchoninic acid (BCA) assay. **g**, Sodium dodecyl sulfate-polyacrylamide gel electrophoresis (SDS-PAGE) showing the presence of all NF components and APA in the AN gel. Data are presented as mean ± s.d. (n = 4). **h**, Western blot analysis of TGF-β1, VEGFA, FGF2, PDGFα and HIF-1α expression in fat, NF and AN gel using GADPH as a loading control. NS at *P* >0.05; *****P*<0.0001 by one-way AVONA and Tukey's post-hoc test.

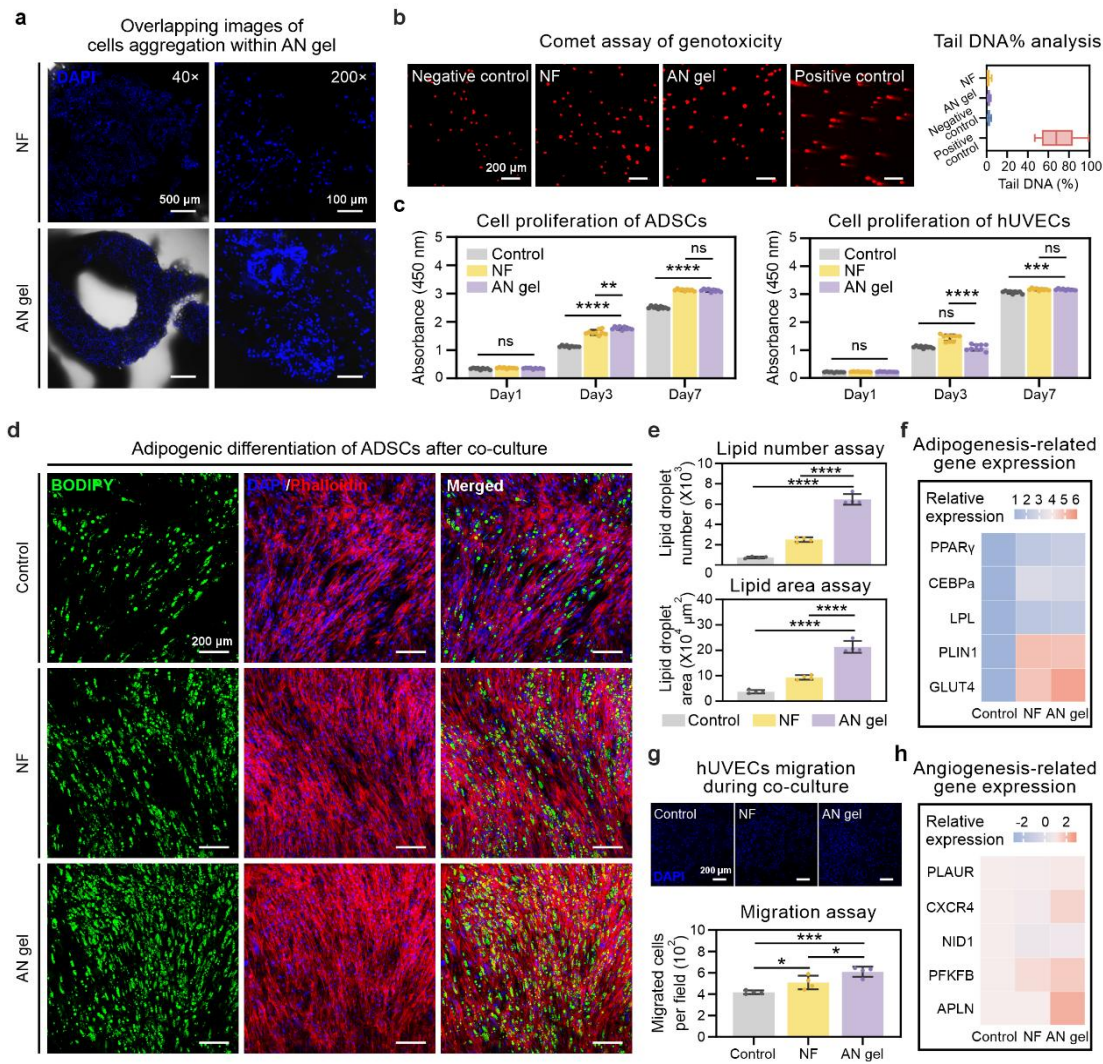

**Fig. 4 | *In vitro* biological validation of the AN gel. a**, Z-stack images showing spatial distribution of nuclei (DAPI, blue) within materials at low and high magnification. **b**, Confocal images (left) of intact or comet-like nuclei (EB, red) in adipose-derived stem cells (ADSCs) co-cultured with materials. DNA damage quantified by tail DNA percentage (right). Each point represents one nucleus (mean ± s.d., n = 50). **c**, Cell viability of ADSCs and human umbilical vein endothelial cells (hUVECs) co-cultured with materials assessed by CCK-8 assay on day 1, 3, and 7 (absorbance at 450 nm). **d**, ADSCs after co-culture and adipogenic induction were stained with BODIPY (lipids, green), phalloidin (F-actin, red), and DAPI (nuclei, blue). Merged and individual channels are shown. **e**, Quantification of lipid droplet number (top) and total lipid area (bottom) per field. Data are presented as mean ± s.d. (n = 4 random regions) **f**, Heatmap shows relative expression of adipogenic-related genes in co-cultured ADSCs, normalized to control. Data represent means of three replicates. **g**, Confocal images (top) of DAPI-stained nuclei showing migrated hUVECs after co-culture. Quantification (bottom) by nuclear count per field. **h**, Heatmap showing relative expression of angiogenesis-related genes in co-cultured hUVECs, normalized to control. Data represent means of three replicates. For **c**,

Data are presented as mean ± s.d. (n = 4). NS at $P > 0.05$; ***$P < 0.001$; ****$P<0.0001$ by two-way AVONA and Tukey's post-hoc test. For **e,g**, Data are presented as mean ± s.d. (n = 4). *$P<0.05$; ***$P < 0.001$; ****$P<0.0001$ by one-way AVONA and Tukey's post-hoc test.

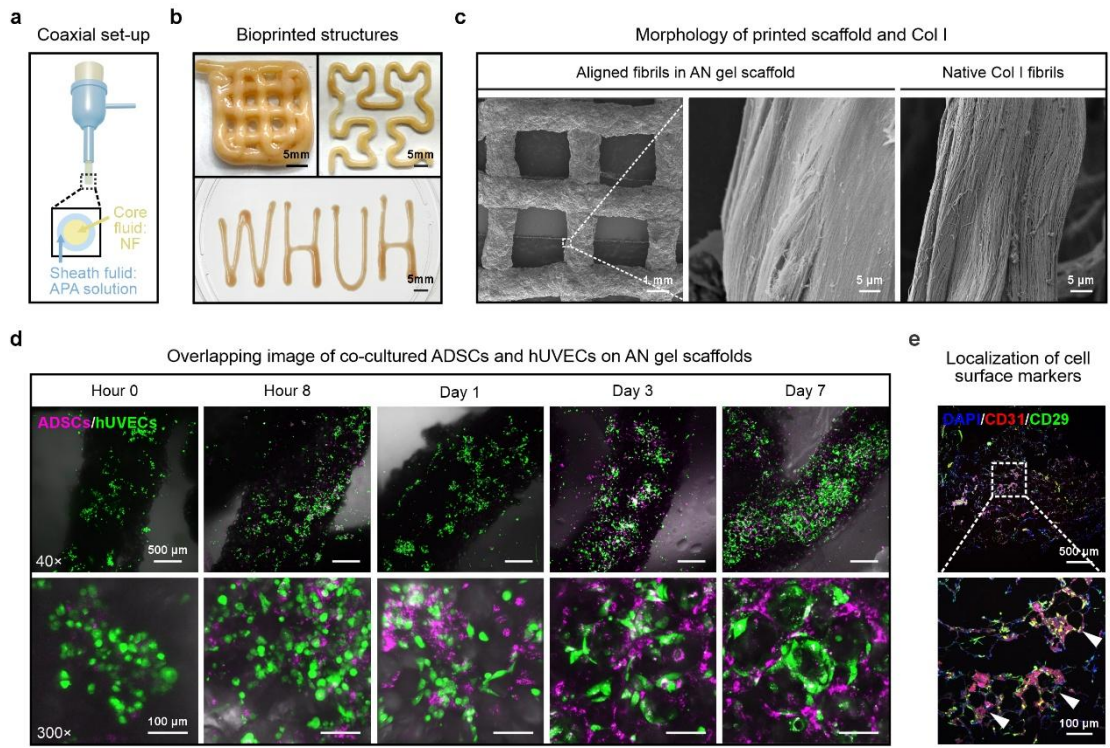

**Fig. 5 | Biofabrication and bioactivity of the AN gel scaffolds. a**, Schematic showing the coaxial printing strategy enabling immediate co-assembly of the AN gel during extrusion, which demonstrating the high printing fidelity and structural complexity of the printed scaffolds (**b**). **c**, Magnified SEM view of scaffold surface showing aligned fibrils form with the printing direction, which are resembled with the native Col I. **d**, Overlapping 3D confocal images tracking the proliferation and spatial interactions between membrane-labeled ADSCs (DiI-labeled, pseudocolored in magenta) and autofluorescent-hUVECs (green) co-cultured on AN gel scaffolds over 7 days. **e**, On day 7, immunofluorescence staining for CD31 (hUVECs marker) and CD29 (ADSCs marker) revealing the cellular organization within the scaffold.

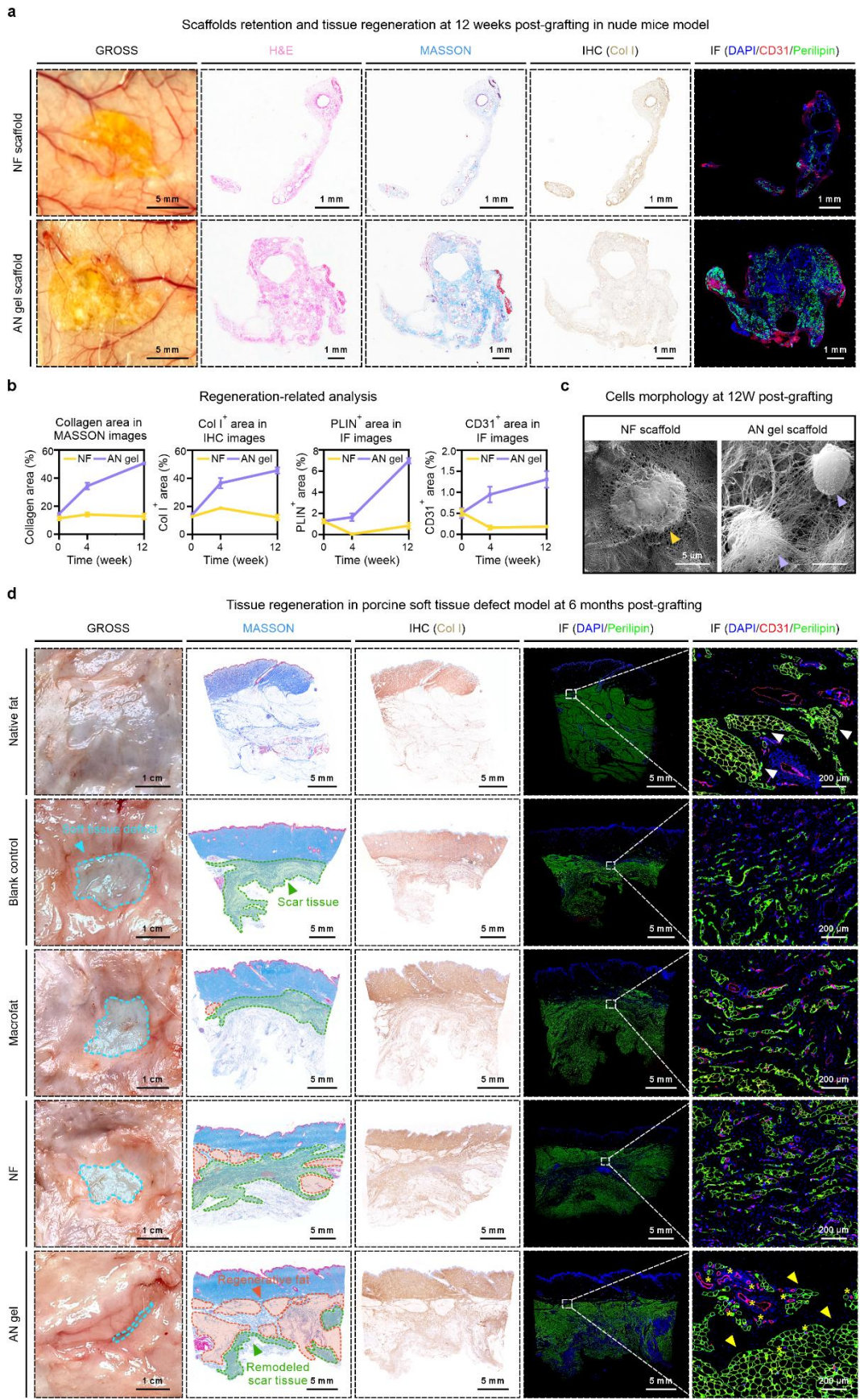

**Fig. 6 | *In vivo* vascular soft tissue regeneration promotion. a**, Representative images of retained scaffolds and regenerated tissues at 4 or 12 weeks post-surgery in the nude mice model.

Histological morphology was visualized by H&E staining, collagen deposition by Masson staining (MASSON), Col I accumulation by IHC staining (Col I), and vascularized adipocyte clusters by IF staining (DAPI/CD31/Perilipin). **b**, Quantification of collagen area (MASSON), Col I-positive area (IHC), and PLIN- and CD31-positive areas (IF). Data are presented as mean ± s.d. (n = 4 random regions). **c**, SEM images of scaffold surfaces after 12 weeks of subcutaneous grafting in nude mice. Yellow arrows indicate atrophic cells, and purple arrows indicate plump cells. **d**, Representative images of regenerated soft tissue at 6 months post-surgery in the soft tissue defect porcine model. Gross appearance (GROSS), collagen deposition (MASSON), Col I accumulation (IHC), and vascularized adipocyte clusters (IF, DAPI/Perilipin at low magnification, DAPI/CD31/Perilipin at high magnification) are shown. Blue overlays indicate defect areas, green overlays indicate scar tissue, and orange overlays indicate regenerated fat. Yellow asterisks mark CD31-positive vessels, and yellow arrows indicate PLIN-positive adipocytes.